\documentclass[prb,showpacs,twocolumn,floatfix]{revtex4}
\usepackage{graphicx}
\begin{document}
\title{Field dependence of magnetic ordering in Kagom\'{e}-staircase compound ${\bf Ni_3V_2O_8}$}

\author{M. Kenzelmann,$^{1,2}$ A. B. Harris,$^3$ A. Aharony,$^4$
O. Entin-Wohlman,$^4$ T.  Yildirim,$^2$ Q. Huang,$^{2}$ S. Park,$^{2}$
G. Lawes,$^5$ C. Broholm,$^{1,2}$ N. Rogado,$^6$ R. J.  Cava,$^6$
K. H.  Kim,$^5$ G. Jorge,$^5$ and A. P. Ramirez$^{5,7}$}

\affiliation{(1) Department of Physics and Astronomy, Johns
Hopkins University, Baltimore, Maryland 21218\\
(2) NIST Center for Neutron Research, National Institute of Standards
and Technology, Gaithersburg, Maryland 20899\\
(3) Department of Physics and Astronomy,
University of Pennsylvania, Philadelphia, PA 19104\\
(4) School of Physics and Astronomy, Raymond and Beverly Sackler Faculty of
Exact Sciences, Tel Aviv University, Tel Aviv 69978, Israel\\
and Department of Physics, Ben Gurion University, Beer Sheva 84105, Israel\\
(5) Los Alamos National Laboratory, Los Alamos, New Mexico 87545\\
(6) Department of Chemistry and
Princeton Materials Institute, Princeton University, Princeton,
New Jersey 08544\\
(7) Bell Labs, Lucent Technologies, 600 Mountain
Avenue, Murray Hill, NJ 07974}
%%% ----------------------------------------------------------------------
\date{\today}

\begin{abstract}
We present powder and single-crystal neutron diffraction and bulk
measurements of the Kagom\'{e}-staircase compound ${\rm Ni_3V_2O_8}$
(NVO) in fields up to $8.5$T applied along the {\bf c}-direction.
(The Kagom\'{e} plane is the {\bf a}-{\bf c} plane.) This system
contains two types of Ni ions, which we call ``spine" and
``cross-tie". Our neutron measurements can be described with the
paramagnetic space group ${\rm Cmca}$ for $T < 15$K and each
observed magnetically ordered phase is characterized by the
appropriate irreducible representation(s). Our zero-field
measurements show that at $T_{\rm PH}=9.1$K NVO undergoes a
transition to an incommensurate order which is dominated by a
longitudinally-modulated structure with the spine spins mainly
parallel to the {\bf a}-axis. Upon further cooling, a transition is
induced at $T_{\rm HL}=6.3$K to an elliptically polarized
incommensurate structure with both spine and cross-tie moments in
the {\bf a}-{\bf b} plane. At $T_{\rm LC}=4$K the system undergoes a
first-order phase transition, below which the magnetic structure is
a commensurate antiferromagnet with the staggered magnetization
primarily along the {\bf a}-axis and a weak ferromagnetic moment
along the {\bf c}-axis. A specific heat peak at $T_{\rm CC'}=2.3$K
indicates an additional transition, which we were however not able
to relate to a change of the magnetic structure. Neutron, specific
heat, and magnetization measurements produce a comprehensive
temperature-field phase diagram. The symmetries of the two
incommensurate magnetic phases are consistent with the observation
that only one phase has a spontaneous ferroelectric polarization.
All the observed magnetic structures are explained theoretically
using a simplified model Hamiltonian, involving competing nearest-
and next-nearest-neighbor exchange interactions, spin anisotropy,
Dzyaloshinskii-Moriya and pseudo-dipolar interactions.
\end{abstract}
 \maketitle

\section{Introduction }

Quantum spin systems with competing interactions can have
highly-degenerate ground-state manifolds with unusual spin
correlations. Small, otherwise insignificant perturbations can then
become decisively important by removing the degeneracy of the
low-energy spin fluctuations, leading to unexpected ground states at
low temperatures. The proximity to quantum phase transitions, which
separate the various ground states, can lead to new types of
instabilities which involve charge and lattice degrees of freedom.
Examples include exotic superconductivity, heavy-fermion conductors
and ferroelectricity.\par

Frustrated low-spin magnets are ideal model systems for the study of
competing quantum phases because they naturally contain competing
interactions in a clearly-defined geometry. An important model
system is the Kagom\'{e} lattice which consists of corner-sharing
triangles of spins with equal antiferromagnetic (AF) coupling
between nearest neighbors. Theoretically it is expected that the
$S=\frac{1}{2}$ Kagom\'{e} lattice does not have long-range order at
zero temperature, but adopts a quantum spin liquid ground
state.\cite{Sachdev,Sindzingre} The most well known
Kagom\'{e}-related magnet is the $S=\frac{3}{2}$ compound ${\rm
SrCr_9Ga_3O_{19}}$.  However, the structure is actually better
described as a Kagom\'{e} bilayer with an intervening triangular
lattice\cite{NEW} and it has a ground spin-glass like ground
state.\cite{BroholmAeppli,RamirezEspinosa} Work on jarosite systems
showed various types of commensurate long-range order stabilized by
inter-layer and Dzyaloshinskii-Moriya (DM)
interactions.\cite{LeeBroholm,YSLee} These results indicate that
Kagom\'{e} related magnets are highly sensitive to relatively weak
interactions, and hence are a likely venue for new and exotic
ordered states.\par

${\rm Ni_3V_2O_8}$ (NVO) is a system of weakly-coupled spin-1
staircase Kagom\'{e} planes containing two types of inequivalent
bonds and magnetic sites. It is thus a variant of the
highly-frustrated pure Kagom\'{e} lattice. However, these deviations
from ideal Kagom\'{e} geometry introduce several new interactions
which determine how the quantum degeneracy and frustration are
resolved.  In addition, it has been found\cite{FERRO} that the
magnetic ordering generates ferroelectricity and these smaller
interactions play a crucial role in this phenomenon and can explain
the microscopic origins of multiferroics.\cite{TOBE}\par

\begin{figure}
\begin{center}
\includegraphics[height=10cm,bbllx=85,bblly=50,bburx=445,
bbury=540,angle=0,clip=]{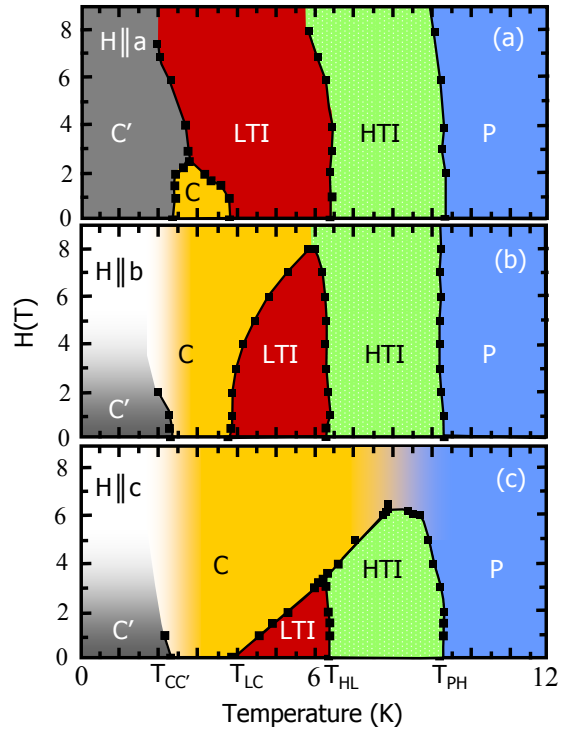} \caption{\label{PHASED} Phase
diagram in $H$-$T$ when ${\bf H}$ is applied along each of the three
crystallographic axes. This phase diagram is based on specific heat
data taken as a function of $T$ at constant $H$
(Ref~\protect\onlinecite{LawesKenzelmann}). The phases are labeled P
for the paramagnetic (magnetically disordered) phase, HTI for the
high-temperature incommensurate phase, LTI for the low-temperature
incommensurate phase, C for the high-temperature canted AF phase,
and C' for the low-temperature canted AF phase. The spin structures
of these phases are described below. The transition temperatures (in
order of decreasing temperature) are denoted $T_{\rm PH}$, $T_{\rm
HL}$, $T_{\rm LC}$, and $T_{\rm CC'}$. For a field along ${\bf c}$,
the C and P phases have the same symmetry, and therefore there is no
sharp transition between them.}
\end{center}
\end{figure}

Magnetic susceptibility and specific heat measurements reveal that
NVO undergoes a series of magnetic phase transitions with
temperature and magnetic field\cite{Rogado,LawesKenzelmann} and in
Fig. \ref{PHASED} we reproduce the phase diagram as found in Ref.
\onlinecite{LawesKenzelmann}, which we refer to as I. There, our
neutron diffraction study at zero field showed that NVO adopts two
different incommensurate phases above $4$K, a mainly longitudinal
incommensurate phase which occurs at higher temperature (which we
therefore call the high-temperature incommensurate, HTI, phase) and
a spiral incommensurate phase which occurs at lower temperature and
which we call the low-temperature incommensurate, or LTI, phase. We
also found evidence of a commensurate canted AF phases below $4$K.
One purpose of this paper is to present a comprehensive review of
neutron diffraction data which enables us to characterize the HTI,
LTI and C phases, but we will not discuss in detail the C' phase. A
second purpose is to show that the phase diagram can be understood
to be the result of competing nearest-neighbor (nn) and
next-nearest-neighbor (nnn) exchange interactions, easy-axis
anisotropy, and anisotropic interactions, both pseudo-dipolar (PD)
as well as DM interactions. It is particularly important to
characterize the magnetic structure of NVO in order to gain an
understanding of the magnetoelectric coupling\cite{FERRO} which
causes the LTI phase to be ferroelectric.\par

The experimental details, data analysis, and physical interpretation
which were only briefly presented in I, are fully explained in this
paper. We performed a zero-field powder and single-crystal neutron
diffraction study to determine the magnetic structures, and we used
group theory to identify the structures that are allowed by symmetry
for the two observed ordering wave-vectors. Further we present the
field dependence of the magnetic structures by monitoring AF and
ferromagnetic (FM) Bragg peaks. Magnetic fields up to $8.5$T were
applied along the crystallographic $\bf c$-direction. We find that
application of a magnetic field along the $\bf c$-axis favors the AF
C phase at the expense of the incommensurate phases. Thus the phase
diagram obtained by our neutron diffraction experiments is
consistent with that obtained from macroscopic measurements, some of
which are presented here and provide additional information on the
microscopic interactions between magnetic ions.\par

Almost all our results can be understood on the basis of theoretical
models that are analyzed in detail in a separate paper,\cite{THEORY}
which we refer to as II. To avoid undue repetition we will here
indicate in qualitative terms how these models explain the data and
we refer the reader to II for quantitative details.\par

Briefly this paper is organized as follows. In Sec. II we summarize
the experimental techniques used in this work. In Sec. III we give
our determination of the crystal crystal which confirms earlier work
by Sauerbrei {\it et al.}\cite{Sauerbrei} Section IV contains the
magnetic structure determinations. Here we give a brief summary of
representation theory, since this forms the basis for almost all the
structure determinations. We also present magnetization and
susceptibility data for magnetic fields applied along the three
crystallographic directions. In Sec. VI we give a theoretical
interpretation of the experimental results. We obtain rough
estimates of many of the microscopic interaction parameters using
the more detailed calculations presented in II. Finally, our results
are summarized in Sec. VI.

\section{Experimental}
Powder samples of NVO were made in a crucible using ${\rm NiO}$
and ${\rm V_2O_5}$ as starting materials.\cite{Rogado} Single
crystals were grown from a ${\rm BaO}$-${\rm V_2O_5}$
flux.\cite{LawesKenzelmann} Neutron diffraction experiments were
performed using a powder sample with a total mass of
$10\;\mathrm{g}$ and a single crystal with a mass of
$0.13\;\mathrm{g}$ oriented with the $(h,k,0)$ or $(h,k,k)$
crystallographic planes in the horizontal scattering plane of the
spectrometers.\par

Powder neutron measurements were performed using the BT-1
high-resolution powder diffractometer at the NIST Center for Neutron
Research, employing Cu (311) and Ge (311) monochromators to produce
a monochromatic neutron beam of wavelength $1.5401$\AA \ and
$2.0775$\AA, respectively. Collimators with horizontal divergences
of $15'$, $20'$, and $7'$ were used before and after the
monochromator, and after the sample, respectively. The intensities
were measured in steps of $0.05^{o}$ in the range of scattering
angles, $2\Theta$, from $3^{o}$ to $168^{o}$. Data were collected at
various temperatures from $1.5$K to $30$K to elucidate the
temperature dependence of the crystal structure. The program GSAS
\cite{GSAS} was used to refine the structural parameters and the
commensurate magnetic structure. Additional diffraction data and
magnetic order parameters were obtained on the BT7 triple axis
spectrometer to explore the magnetic scattering in more detail. For
these measurements, a pyrolytic graphite PG(002) double
monochromator was employed at a wavelength of $2.47$\AA, with $40'$
collimation after the sample and no analyzer.\par

\begin{table}
\begin{scriptsize}
\begin{tabular}{c|c|c|c}
\hline\hline
   &  & $T= 15$K & $T=1.5$K  \\ \hline
a (\AA)  &   &  5.92179(3) & 5.92197(3) \\
b (\AA)  &   &  11.37105(7) & 11.37213(7) \\
c (\AA)  &   &  8.22638(5) & 8.22495(5) \\
${\rm Ni_s}$  &  y  & 0.1299(1) &  0.1299(1)\\
       &  B(\AA$^2$)   & 0.26(1) & 0.24(1)\\
%       & $M_x(\mu_B)$     & &    1.64(4)\\
${\rm Ni_c}$  & B(\AA$^2$) &      0.24(2) & 0.28(2)\\
%       & $M_x(\mu_B)$ & & 0\\
V  &  y &  0.3762 & 0.3762 \\
   &  z &  0.1196 & 0.1196 \\
      &  B(\AA$^2$) &      0.24  &  0.24\\
${\rm O_1}$  &  y  & 0.2482(2) &  0.2482(2) \\
      & z  & 0.2308(2) &  0.2309(2) \\
      &  B(\AA$^2$) &       0.31(2) & 0.30(2) \\
${\rm O_2}$  &  y  & 0.0012(2) &  0.0008(2) \\
      & z  & 0.2443(2) & 0.2447(2) \\
      &  B(\AA$^2$) &       0.32(2) & 0.31(2) \\
${\rm O_3}$ &   x  &  0.2656(2) &  0.2660(2) \\
     & y  & 0.1190(1)   & 0.1191(1)\\
     & z  &  0.00039(8)  & 0.00029(8)\\
     & B(\AA$^2$) & 0.34(2)& 0.30(2)\\
$R_{\rm p}$ (\%) & & 3.80  &  3.97\\
$R_{\rm wp}$(\%) & & 4.69 &   4.78\\
$\chi^2$  & & 1.245 &  1.294\\
& & \multicolumn{2} {c} {distances in $\AA$} \\
${\rm Ni_s}$-${\rm O_2}$ &  2  & 2.010(2)  &  2.013(2) \\
${\rm Ni_s}$-${\rm O_3}$ &  4  & 2.075(2)  &  2.078(1) \\
${\rm Ni_c}$-${\rm O_1}$ &  2  & 2.006(2)  &  2.006(2) \\
${\rm Ni_c}$-${\rm O_2}$ &  2  & 2.083(1)  &  2.085(1) \\
${\rm Ni_c}$-${\rm O_3}$ &  2  & 2.0599(7) &  2.0597(7)\\
${\rm Ni_s}$-${\rm Ni_c}$ & 4  & 2.9330(6) &  2.9331(6)\\
${\rm Ni_c}$-${\rm Ni_c}$ & 2  & 2.96089(2)& 2.96098(2)\\
\hline\hline
\end{tabular}
\caption{\label{Table1}Structural parameters and selected
interatomic distances for NVO, measured using the BT1 spectrometer
with the Ge (311) monochromator and $\lambda$=$2.0775$\AA. Space
group: Cmca (No. 64 in Ref. \protect{\onlinecite{HAHN}}). Atomic
positions (expressed as fractions of $a$, $b$, and $c$): ${\rm
Ni_s}$: 8e (notation as in Ref. \protect{\onlinecite{HAHN}})
($\frac{1}{4}$, y, $\frac{1}{4}$); ${\rm Ni_c}$: 4a (0 0 0 ); V: 8f
(0 y z ); ${\rm O_1}$: 8f (0 y z ); ${\rm O_2}$: 8f (0 y z ); ${\rm
O_3}$: 16g (x y z ). $B_i \equiv 8 \pi^2 \langle u_i^2 \rangle$,
where $u_i$ is the displacement of atom $i$ from its equilibrium
position and $\langle ... \rangle$ indicates a thermal average. Also
$R_{\rm p}$=$\sum^n_i |I^i_o - I^i_c| / \sum_i I^i_o$ where $I^i_o$
and $I^i_c$ are the $n$ observed and calculated intensities,
respectively. $R_{\rm wp}$=$\sqrt{\sum w_i (I^i_o - I^i_c)^2/ \sum
w_i (I^i_o)^2 }$ where the weight is given by $w_i$=$1/\sigma^2_i$.
The sum of least-squares is given by $\chi^2$=$\sum^n_i w_i (|F^i_o|
- |F^i_c|)^2/(n - m)$, where $m$ is the number of fitted variables.}
\end{scriptsize}
\end{table}

The single-crystal neutron scattering measurements were performed
with the thermal-neutron triple-axis spectrometers BT7 and BT9, and
with the cold-neutron triple-axis spectrometer SPINS. The BT7
experiment was performed with 60' collimation after the sample, a
PG(002) analyzer to reflect $13.408\;\mathrm{meV}$ neutrons and 180'
beam divergence after the analyzer. The BT9 diffraction measurements
were performed with an incident energy of $30.5\;\mathrm{meV}$ and
40'-40' collimation around the sample. The $H$-$T$ magnetic phase
diagram was determined using SPINS with an incident energy of
$5\;\mathrm{meV}$, a Be filter before the sample and 80'-80'
collimation around the sample. The SPINS diffraction patterns were
collected with 80'-80' collimation around the sample, a PG(004)
monochromator combined with a graphite filter in the incident beam
and a flat PG(002) analyzer set for $14.7\;\mathrm{meV}$. The beam
divergence between analyzer and detector was 240'.

\section{Nuclear structure}

\subsection{Experimental determination of structure}
The NVO structure refinement from BT1 neutron powder diffraction
data was carried out successfully using the previously reported
structural parameters\cite{Sauerbrei} as initial values. In
agreement with previous studies, we found the structure to have {\it
Cmca} symmetry (space group No. 64 in the International Tables for
Crystallography\cite{HAHN}). No structural transition was detected
for 1.5K$<T<$300K. The structural parameters and selected
interatomic distances at two temperatures are given in
Table~\ref{Table1}. The symmetry elements of the \textit{Cmca} space
group of NVO are given in Table \ref{TablePUC}. Because V has a low
coherent neutron scattering cross-section, the atomic positions and
temperature factors of the vanadium ions were fixed to the values
obtained by Xray diffraction.\cite{Sauerbrei}

\begin{table}
\begin{tabular} {||c | c ||}
\hline
${\rm E}{\bf r} =(x,y,z)$
& $2_c{\bf r} =(\overline x, \overline y + 1/2, z + 1/2)$ \\
 $2_b {\bf r} = (\overline x, y+1/2, \overline z +1/2)$
& $2_a {\bf r} = (x, \overline y , \overline z)$ \\
${\cal I} {\bf r} =(\overline x,\overline y,\overline z)$
&$m_{ab}{\bf r} =(x, y + 1/2, \overline z + 1/2)$ \\
$m_{ac}{\bf r} = (x, \overline y+1/2, z +1/2)$
&$m_{bc}{\bf r} = (\overline x, y , z)$ \\
\hline \end{tabular}
\caption{\label{TablePUC}General positions within the primitive unit cell for
Cmca which describe the symmetry operations of this space group.
$2_\alpha$ is a two-fold rotation (or screw) axis and
$m_{\alpha \beta}$ is a mirror (or glide) $\alpha-\beta$ plane.}
\end{table}

\begin{figure}
  \includegraphics[height=10cm,bbllx=185,bblly=261,bburx=410,
  bbury=540,angle=0,clip=]{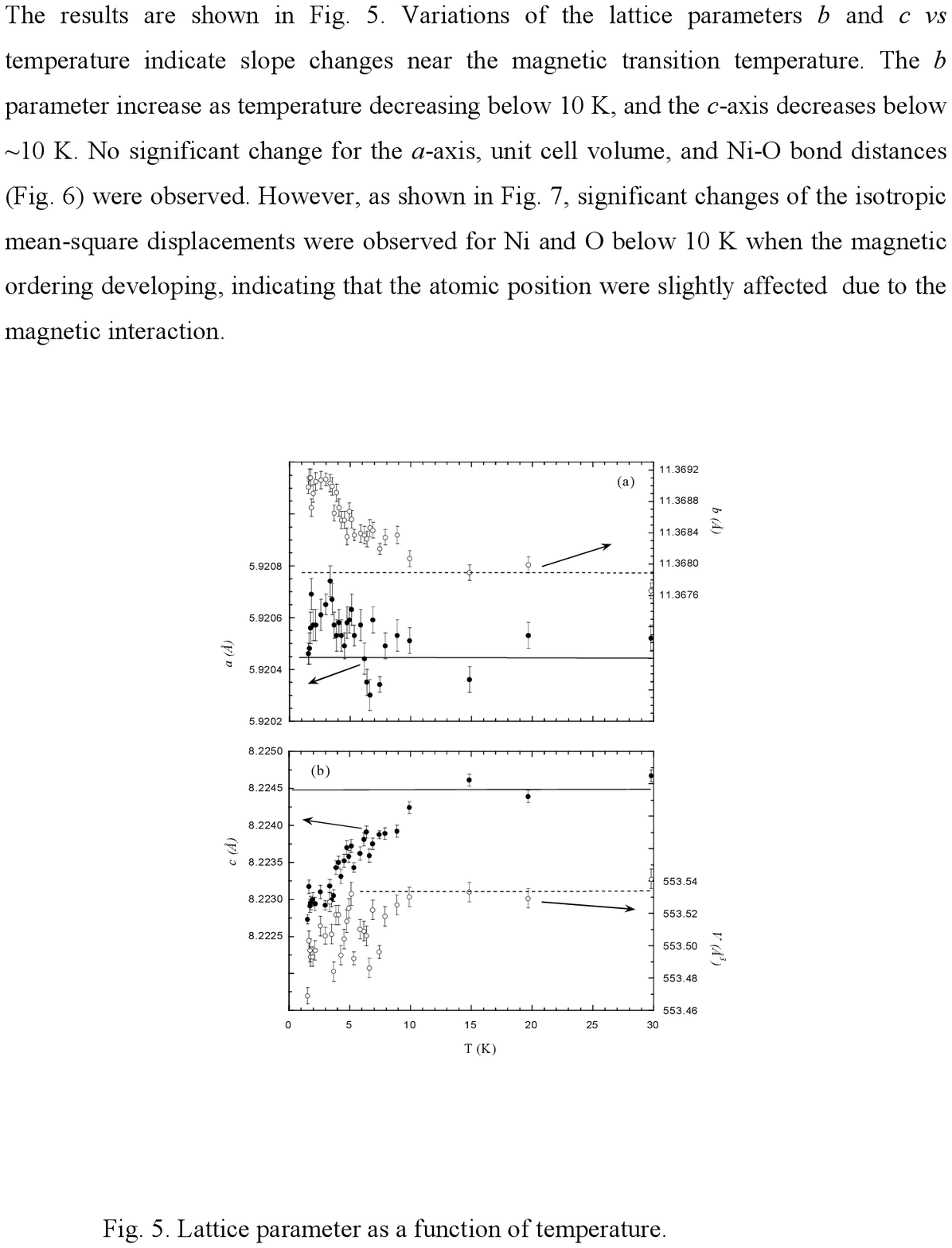}
\caption{\label{Figlattice} Lattice parameters $a$, $b$, and $c$ as
a function of temperature, determined from neutron powder
diffraction using BT1 with a Cu (311) monochromator and a wavelength
1.5401\AA.  Also shown is the temperature dependence of the unit
cell volume.}
\end{figure}

To investigate the effect of magnetic ordering on the chemical
structure, a series of powder patterns were taken below $10$K. No
significant change for the {\bf a}-axis, unit cell volume, and Ni-O
bond distances were observed, as shown in Fig.~\ref{Figlattice} and
Table \ref{Table1}. However, reducing the temperature below $10$K
leads to an increasing $b$ lattice parameter, while the $c$ lattice
parameter  decreases without a change in space group symmetry.\par

\begin{figure} [ht]
\includegraphics[scale=0.4,angle=0]{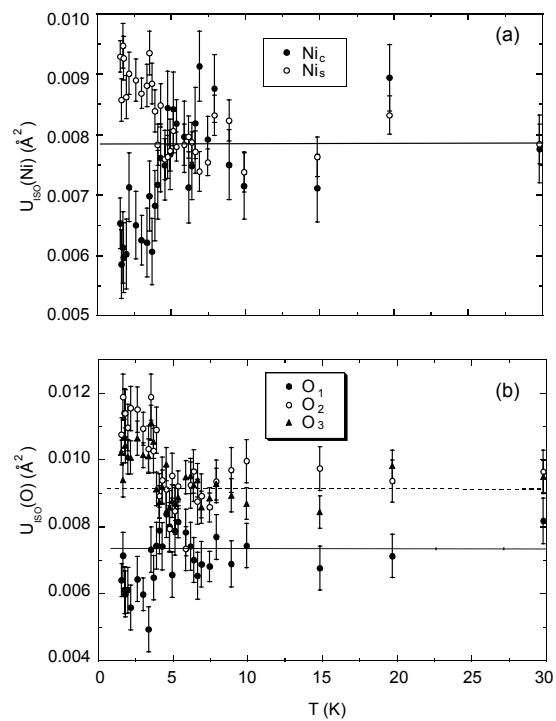}
%\includegraphics[width=2.5in,bbllx8=123,bblly=276,bburx=430,
%bbury=685,angle=0,clip=]{MKFig3v4.eps}
\caption{Isotropic mean-square displacements of Ni and O as a
function of temperature, obtained from the same powder spectra
mentioned in Fig.~\protect\ref{Figlattice}.}
\label{Figmeandisp}
\end{figure}

The strongest temperature dependence is associated with the
isotropic mean-square displacements for Ni and O, which change
significantly with the onset of commensurate order at $4$K
(Fig.~\ref{Figmeandisp}). This may indicate a weak coupling of the
magnetic and the chemical lattice. However, we did not observe a
crystal distortion with the given neutron diffraction resolution.
Higher-resolution xray or neutron diffraction would be needed to
look for possible space group symmetry breaking associated with the
onset of commensurate order.\par

\subsection{Structural properties}
Here we note some general features of the structure. The structure
of NVO consists of ${\rm Kagom\acute{e}}$ layers of edge-sharing
${\rm NiO_6}$ octahedra. The layers are separated by nonmagnetic
${\rm V^{5+}O_{4}}$ tetrahedra. There are two inequivalent
crystallographic sites for Ni, denoted as ${\rm Ni_s}$ and ${\rm
Ni_c}$. (We will refer to these as ``spine" and ``cross-tie" sites,
respectively.) At $15$K, the average ${\rm Ni^{2+}-Ni^{2+}}$
distance within the layers is $d_1=2.94$\AA, while the inter-layer
distance is $d_2=5.69$\AA. Based on the relatively large inter-layer
to intra-layer ratio, $d_2/d_1 = 1.9$, a strong two-dimensional
magnetic character may be expected in this compound, with the
magnetism dominated by intra-layer ${\rm Ni^{2+}-Ni^{2+}}$ exchange
interactions. Unlike previously studied ${\rm Kagom\acute{e}}$
lattice-based materials, which have planar magnetic layers, the Ni-O
layers in NVO are buckled, resulting in the ``${\rm
Kagom\acute{e}}$-staircase" structure. The symmetry of the
superexchange interaction mediated by O ions shows that there are
two inequivalent superexchange paths between neighboring ${\rm
Ni^{2+}}$ ions within ${\rm Kagom\acute{e}}$ planes. In particular,
there is a superexchange path between ${\rm Ni_s}$ positions along
the crystallographic {\bf a}-direction which is different from nn
interactions between Ni ions on neighboring ${\rm Ni_s}$ and ${\rm
Ni_c}$ positions.\par

\begin{figure}[ht]
\begin{center}
  \includegraphics[width=8cm]{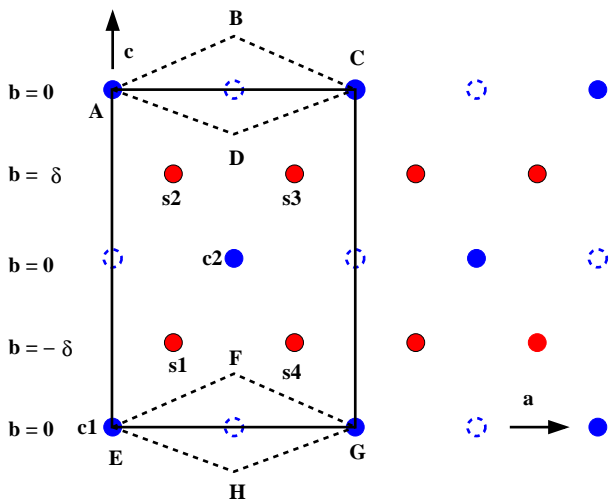}
  \caption{Ni sites in the primitive unit cell, whose vertices are
A, B, \dots H  and is defined by the vectors ${\bf v}_1 =
(a/2)\hat a + (b/2)\hat b$, ${\bf v}_2 = (a/2)\hat a - (b/2)\hat
b$, and ${\bf v}_3=c \hat c$. The ``cross-tie" sites
(on-line=blue) c1 and c2 lie in a plane with $b=0$. The ``spine"
sites (on-line=red) are labelled s1, s2, s3, and s4 and they may
be visualized as forming chains parallel to the ${\bf a}$-axis.
These chains are in the buckled plane with $b= \pm \delta$, where
$\delta = 0.13 b$ as is indicated.  Cross-tie sites in adjacent planes
(displaced by $(\pm b/2)\hat b$) are indicated by open circles.
Spine sites in adjacent planes are located directly above (or
below) the sites in the plane shown.}
\label{STRUCTURE}
\end{center}
\end{figure}

The point group is Abelian and its three generators are spatial
inversion (${\cal I}$), and the two-fold rotations $R_a$ and $R_b$
about the $\bf a$ and $\bf b$ axes respectively. The space group
operations are ${\cal I}$ and $R_a$ referred to the origin at a
cross-tie site, such as site c1 in Fig. \ref{STRUCTURE}, and $R_b$
about an axis passing through a spine site such as site s1 in Fig.
\ref{STRUCTURE}. The operations give rise to mirror planes
perpendicular to the ${\bf a}$-axis passing through a cross-tie site
and a two-fold screw axis about a $z$-axis passing through a spine
site such as site s1 with a translation of $c/2$ along ${\bf c}$.
There is a glide plane perpendicular to the $\bf c$-axis passing
through a chain of spine sites with a translation $a/2$ along ${\bf
a}$.  The result of these symmetry operations is that all the Ni
spine sites are related by symmetry and all the Ni cross-tie sites
are similarly related by symmetry. Our convention for the coordinate
axes is as follows: the spines lie along the (100), or {\bf a}-axis
(denoted sometimes the $x$-axis), the basal plane includes this axis
and the (001), or {\bf c}-axis (or sometimes the $z$ axis), and the
axis perpendicular to this plane is denoted (010) or the {\bf
b}-axis (or sometimes the $y$-axis). The two types of Ni sites are
shown in Fig. \ref{STRUCTURE} and their coordinates are given in
Table \ref{NiLattice}.

\begin{table}
\begin{tabular}{cccc}\hline\hline
&${\bf r}_s^1$  &=&  $(0.25, -0.13,0.25)$    \\
&${\bf r}_s^2$  &=&  $(0.25, 0.13, 0.75)$   \\
&${\bf r}_s^3$  &=&  $(0.75, 0.13, 0.75)$   \\
&${\bf r}_s^4$  &=&  $(0.75, -0.13, 0.25)$  \\
&${\bf r}_c^1$  &=&  $(0,    0,    0)$      \\
&${\bf r}_c^2$  &=&  $(0.5,  0,    0.5)$    \\
\hline\hline
\end{tabular}
\caption{\label{NiLattice} Positions of ${\rm Ni^{2+}}$ carrying
$S$=$1$ within the primitive unit cell illustrated in Fig.
\protect{\ref{STRUCTURE}}. Each component is expressed as a fraction
of the respective lattice constant, so ${\bf r}_s^1 = 0.25{\bf a}
-0.13{\bf b} 0.25{\bf c}$. Lattice positions ${\bf r}_s^n$ are spine
sites and ${\bf r}_c^n$ are the cross-tie sites. NVO orders in space
group ${\rm Cmca}$, so there are six more atoms in the orthorhombic
unit cell which are obtained by a translation of lattice ${\bf r}_1$
through ${\bf r}_6$ by $(0.5,0.5,0)$.}
\end{table}

The spine and cross-tie sites have different local symmetry and
will be seen to have very different magnetic properties. In the
presence of AF ordering of the spine sites, the cross-tie sites
are frustrated if one assumes isotropic AF Heisenberg
interactions; the sum of the moments of the four spine neighbors
of each cross-tie spin vanishes. In this regard, this system is
reminiscent of ${\rm Sr_2Cu_3O_4Cl_2}$ \cite{ChouAharony} and of
various ``ladder" systems which have recently been
studied.\cite{KiryukhinBirgeneau}\par

The above structure has several implications for the magnetic
interactions. As explained in Ref.~\onlinecite{Rogado}, the
leading nn Ni-Ni magnetic coupling arises via superexchange
interactions, mediated by two Ni-O-Ni bonds. For a pair of nn
spine spins, the angles of these two bonds are 90.4$^{\rm o}$ and
95.0$^{\rm o}$. For a spine-cross-tie pair, these angles are
90.3$^{\rm o}$ and 91.5$^{\rm o}$. For the similar case of Cu-O-Cu
bonds, it has been shown that when these angles are close to
90$^{\rm o}$ then the resulting exchange energy is
small,\cite{Tornow} and may even change its sign from FM to AF (as
the angle decreases from 90$^{\rm o}$). Since both Ni and Cu
involve $d$-holes in the high $e_g$ states (within the oxygen
octahedron surrounding Ni or Cu), we expect similar results to
apply for the Ni case. Accordingly, we do not necessarily expect
that nn interactions dominate second neighbor interactions.
Similar calculations for the related nnn coupling, via Cu-O-O-Cu,
gave AF interactions.\cite{Tornow} Similar Ni-O-O-Ni interactions
could compete with the nn interactions, and give rise to
incommensurate structures, as explained below.\par

As a result of the crystal symmetries, there is a limited number
of independent nn interaction matrices. If we write the
interaction between spine spins ${\bf S}(i)$ and ${\bf S}(j)$ as
\begin{eqnarray}
{\cal H}_{ij} &=& \sum_{\alpha \beta} M_{\alpha \beta} (i,j)
S_\alpha (i) S_\beta (j)\, ,
\end{eqnarray}
where $\alpha=a,b,c$ is a component label, then once we have
specified $M_{\alpha \beta} (s1,s4)$ (in the notation of Fig.
\ref{STRUCTURE}), we can express the interaction matrices for all
other nn pairs of spine sites in terms of $M_{\alpha
\beta}(s1,s4)$. Similarly, we only need to specify a single
interaction matrix, e. g. $M_{\alpha \beta} (s1,c2)$, for nn pairs
of spine and cross-tie sites or for interactions between nnn in
the same spine. Symmetry also places some restrictions on the form
of $M_{\alpha \beta} (s1,s4)$. These results are obtained in II.

\section{Magnetic structure}

\subsection{Representation theory}
Since group theoretical concepts are central to our analysis, we
shall here summarize the results which we will invoke. Additional
details are available in II and in Appendix A. In general, Landau
theory indicates that the free energy $F$ in the disordered phase is
dominantly a quadratic form in the spin amplitudes $S_\alpha ({\bf
r})$ at site ${\bf r}$. Thus
\begin{eqnarray}
F &=& {1 \over 2} \sum_{{\bf r}, {\bf r}', \alpha \beta}
v_{\alpha \beta} ({\bf r}, {\bf r}') S_\alpha({\bf r})
S_\beta ({\bf r}') \  .
\end{eqnarray}
In view of translational invariance this free energy can be written
in terms of Fourier amplitudes as
\begin{eqnarray}
F &=& {1 \over 2} \sum_{{\bf Q}, \tau, \tau', \alpha , \beta}
v_{\alpha \tau; \beta \tau'}({\bf Q}) S_\alpha({\bf Q}, \tau)
S_\beta ({\bf -Q}, \tau') \ , \label{FPEQ}
\end{eqnarray} where ${\bf Q}$ is the wavevector\cite{CONV} and
$\tau$ labels sites within the unit cell. Here the Fourier
amplitudes are given by
\begin{eqnarray}
S_\alpha ({\bf Q},\tau) = N_{uc}^{-1} \sum_{R}
S_\alpha ({\bf R} + {\bf r}_\tau)
e^{i{\bf Q} \cdot ({\bf R}+{\bf r}_\tau)} \ ,
\end{eqnarray}
where ${\bf r}_\tau$ is the position of the $\tau$th site within the
unit cell and the sum over ${\bf R}$ is over translation vectors for
a system of $N_{uc}$ unit cells. As our data indicate, the ordering
transition is a continuous one which is signaled by one of the
eigenvalues of the quadratic free energy passing through zero as the
temperature is reduced. This condition will be satisfied by some
wavevector, or more precisely, by the star of some wavevector ${\bf
q}$, which is in the first Brillouin zone of the primitive lattice.
(This is usually called ``wavevector selection.") The {\it critical
eigenvector}, {\it i. e.} the eigenvector associated with this
instability, will indicate the pattern of spin components within the
unit cell which forms the ordered phase. In view of the symmetry of
the paramagnetic crystal, which dictates the form of Eq.
(\ref{FPEQ}), we see that this eigenvector has to transform
according to one of the representations of the symmetry group which
leaves the wavevector ${\bf q}$ invariant.\cite{Heine} (This group
is usually called ``the little group.") The assumption that the
instability (toward the condensation of long range order) involves
only a single representation is based on the assertion that there
can be no accidental degeneracy (which would correspond to a higher
order multicritical point). Here we are interested in the
representations of two types of wavevectors, namely zero wavevector
(in which antiferromagnetism arises because of AF interactions
within the unit cell) and an incommensurate wavevector $(q,0,0)$ at
some nonspecial point on the $x$-axis. Because of the magnetic
structure of the unit cell one has to be careful in relating the
selected wavevector $(q,0,0)$ to the physically relevant quantity,
namely the Fourier coefficient of the magnetic moment. The
representations for the wavevector $(q,0,0)$ are described in an
Appendix. Within a given representation $\Gamma$ one has the
parameters $m_{sa}^{(\Gamma)}$, $m_{sb}^{(\Gamma)}$ and
$m_{sc}^{(\Gamma)}$ which completely fix the $\bf a$, $\bf b$, and
$\bf c$ components, respectively, of the Fourier amplitudes
$S_\alpha ({\bf Q}, {\rm s}n)$ of all the spine spins within the
unit cell and the parameters $m_{ca}^{(\Gamma)}$,
$m_{cb}^{(\Gamma)}$, and $m_{cc}^{(\Gamma)}$ which similarly
completely fix the Fourier amplitudes $S_\alpha({\bf Q},{\rm c}n$)
of all the cross-tie spins within the unit cell. For some
representations some of these amplitudes may not be allowed to be
nonzero. For an incommensurate wavevector these parameters are
complex-valued, although, of course, the resulting spin components
must be real, because we should invoke both $\Gamma$ associated with
${\bf q}$ and $\Gamma^*$ associated with $-{\bf q}$.

Happily, as the above discussion implicitly assumes, the situation
is quite simple in that for many systems, such as NVO, all the
representations are one dimensional. What that means is that under
any group operation, the allowed eigenvectors transform either into
themselves or into a phase factor of unit magnitude times
themselves. To summarize the results of Appendix A, if ${\cal O}_p$
is an operation that leaves the incommensurate nonzero wavevector
invariant, then we may write
\begin{eqnarray}
{\cal O}_p m_{ g \alpha}^{(\Gamma)} = \xi_p(\Gamma)
m_{g \alpha}^{(\Gamma)} \ ,
\end{eqnarray}
where $g$ assumes the values s for spine and c for cross-tie,
$\alpha = a, b$, or $c$, and $\xi_p(\Gamma)$ is the character for
the symmetry operation ${\cal O}_p$ in the representation $\Gamma$.
For zero wavevector (relevant for the AF phases) these characters
are given in Table \ref{TableIrreducibleRepCom} and for the
incommensurate phases they are given in Table
\ref{TableIrreducibleRepInCom}.

The discussion up to now took account only of those operations which
leave the wavevector invariant. However, the free energy must be
invariant under all the symmetry operations of the paramagnetic
phase.\cite{IED,LL} Thus far the symmetry properties we have
discussed apply to any crystal whose paramagnetic space group is
Cmca. Now we discuss the consequences of restricting the magnetic
moments to the spine and cross-tie sites which have higher site
symmetry than an arbitrary lattice site and in particular we will
study the incommensurate phases. We consider the effect of spatial
inversion on the spin wavefunctions. To illustrate the concepts we
assume a wavevector $q$ (in units of $2 \pi/a$) along the ${\bf
a}$-axis and consider a spin configuration which transforms
according to $\Gamma_4$ and which therefore has the components
$\psi_4$ given in Table \ref{TableIrreducibleRepInCom}. Since the
magnetic moment is an axial vector, spatial inversion ${\cal I}$
takes the moment into itself but moves it to the spatially inverted
lattice site. Let us consider the spin state at spine site \#3 in
the unit cell at ${\bf R}\equiv (X,Y,Z)$. Before applying spatial
inversion the spin vector at that site is
\begin{eqnarray}
&& {\bf S}_3({\bf R};\Gamma_4) = [m_{sa} \hat {\bf a} - m_{sb} \hat
{\bf b} + m_{sc} \hat {\bf c}] e^{2 \pi i q(X+x_{s3})/a} \nonumber \\
&& \ \ \ \ + [m_{sa} \hat {\bf a} - m_{sb} \hat {\bf b} + m_{sc} \hat
{\bf c}]^* e^{-2 \pi i q(X+x_{s3})/a} \ , \label{BEFORE}
\end{eqnarray} where $x_{s3}$ is the $x$-coordinate of spine site
s3 within the unit cell and the representation label is implicit.
After inversion (indicated by a prime) the spin at this site will
be that which before inversion was at $-{\bf R}- x_{s3}\hat {\bf
a}$, which is a site of sublattice \#1. Thus
\begin{eqnarray}
&& {\bf S}_3({\bf R};\Gamma_4)^\prime = [m_{sa} \hat {\bf a} +
m_{sb} \hat {\bf b} + m_{sc} \hat {\bf c}] e^{-2\pi i q(X+x_{s3})/a} \nonumber \\
&& \ \ \ \ + [m_{sa} \hat {\bf a} + m_{sb} \hat {\bf b} + m_{sc}
\hat {\bf c}]^* e^{2 \pi i q(X+x_{s3})/a} \ . \label{AFTER}
\end{eqnarray} By comparing Eqs. (\ref{BEFORE}) and (\ref{AFTER}) we
see that for $\alpha=a$ or $\alpha=c$ we have that
\begin{eqnarray}
{\cal I} [ m_{s \alpha} ] = [m_{s \alpha}]^* \ ,
\label{SYMEQ} \end{eqnarray}
and for the $b$-component we have
\begin{eqnarray}
{\cal I} m_{sb} = -m_{sb}^* \ .
\label{MINUS} \end{eqnarray}
One can check that the spin components of the other spine sites
transform this same way.  Furthermore, this type of analysis
indicates that all the coordinates of Table \ref{TableIrreducibleRepInCom}
for the cross-tie sites obey Eq. (\ref{SYMEQ}).

Accordingly, for this irreducible representation (irrep) we now
introduce {\it symmetry-adapted coordinates} $\tilde m_{g\alpha}$
which obey Eq. (\ref{SYMEQ}). We write
\begin{eqnarray}
\tilde m_{g\alpha} = m_{g\alpha} \ ,
\end{eqnarray}
except for  $g=s$ and $\alpha=b$.  For this case, to transform coordinates
so that Eq. (\ref{MINUS}) is transformed into the desired form of
Eq. (\ref{SYMEQ}), we write
\begin{eqnarray}
\tilde m_{sb} = im_{sb} \ . \label{III} \end{eqnarray} For the
other irreps the analysis is similar, but the components which
have to transform as in Eq. (\ref{III}) may be different. The
complex-valued symmetry adapted coordinates which transform
according to each of the irreps are collected in Table \ref{SYMAD}
and all of these are constructed to obey Eq. (\ref{SYMEQ}).

When we assume the condensation of a single irrep, $\Gamma$, the Landau
expansion in terms of the above symmetry adapted coordinates is of the form
\begin{eqnarray}
F &=& {1 \over 2} \sum_{g g'\alpha \beta} v_{gg'}^{\alpha \beta}
\tilde m_{g\alpha}^* \tilde m_{g'\beta} \ ,
\end{eqnarray}
where $\tilde m_{g \alpha}$ is shorthand for $\tilde m_{g \alpha}^\Gamma$
and the reality of $F$ requires that
\begin{eqnarray}
v_{gg'}^{\alpha \beta} = {v_{g'g}^{\beta \alpha}}^* \ .
\label{HERMIT} \end{eqnarray} Because these coordinates transform
according to a single one-dimensional irrep we know that this
expression for the free energy is indeed invariant under the
operations of the little group. But the free energy is also
invariant under spatial inversion. This additional invariance
provides addition information. This situation has been reviewed
recently by J. Schweizer\cite{JS}, but the approach we use below may
be simpler in the present case. Here, because of the special
transformation property of Eq. (\ref{SYMEQ}) we have that
\begin{eqnarray}
F &=& {\cal I} F = {1 \over 2} \sum_{gg'\alpha \beta} v_{gg'}^{\alpha \beta}
({\cal I}\tilde m_{g\alpha})^* {\cal I} \tilde m_{g'\beta} \nonumber \\ &=&
{1 \over 2} \sum_{gg'\alpha \beta} v_{gg'}^{\alpha \beta}
(\tilde m_{g\alpha}) (\tilde m_{g'\beta})^* \ .
\end{eqnarray}
Taking account of Eq. (\ref{HERMIT}) this implies that
\begin{eqnarray}
v_{g'g}^{\beta \alpha} = v_{gg'}^{\alpha\beta}
={v_{g'g}^{\beta \alpha}}^* \ .
\end{eqnarray}In other words, for the present symmetry, the
coefficients in the quadratic free energy (expressed in terms of
symmetry-adapted coordinates) are all real-valued! What this means
is that the eigenvector of the quadratic form can be expressed as a
single overall complex phase factor times a vector with real-valued
components. This condition ensures that all the amplitudes which
make up the spin eigenvector have the same phase.

The above discussion incorporates several implicit assumptions. For
instance, it is assumed that the magnetic moment is truly localized
on the high symmetry Ni sites, whereas in reality this moment is
spread out around these sites. Also, in principle even the
``nonmagnetic" oxygen atoms will have a small induced magnetic
moment. In addition one can consider the effects of quartic and
higher-order terms in the Landau free energy as well as fluctuations
not included by mean field theory. These corrections are analyzed in
II.

\subsection{Domains}

We now discuss the effect of domains for the case of zero applied
field. In this case if we assume a single spin eigenfunction,
$\Psi_1$, then, because we condense order out of the paramagnetic
phase, we expect to have all eight states of the type ${\cal O}_i
\Psi_1$, where ${\cal O}_i$ is one of the eight symmetry elements of
the paramagnetic space group. The simplest way to take account of
these possibly different structures is to associate with a given
scattering vector ${\bf Q}$, the intensity averaged over the set of
eight wavevectors ${\cal O}_i {\bf Q}$. These eight states will not,
in general, be distinct, but this is a simple automatic way to take
domains into account. Indeed, for the HTI phase, these eight states
will generate four times the state $\Psi_1$ and four times the state
$-\Psi_1$. Since these two configurations both give the same neutron
scattering signal, there is no effect due to domains, as is the case
for two sublattice antiferromagnets. The result is less trivial when
we have simultaneous appearance of two different representations as
in the LTI phase, where we have $\Gamma_4$ and $\Gamma_1$, where the
Fourier components can either be added or subtracted from each
other. Within the accuracy of the experiment we could not
distinguish whether or not both such domains occur simultaneously in
NVO.

This procedure can be extended to nonzero magnetic field ${\bf
H}$. If the field is large enough (or has been obtained by
reducing the field from a large initial value), one would assume
that all domains obtained by applying the operations ${\it o}_i$
of the subgroup of the space group {\it which leaves ${\bf H}$
invariant} occur with equal probability. Then one associates with
a given scattering vector ${\bf Q}$ the intensity averaged over
the set of wavevectors ${\it o}_i {\bf Q}$.

\subsection{Magnetic order at zero field}

\begin{figure}[ht]
\begin{center}
\includegraphics[height=12cm,bbllx=100,bblly=110,bburx=470,
bbury=750,angle=0,clip=]{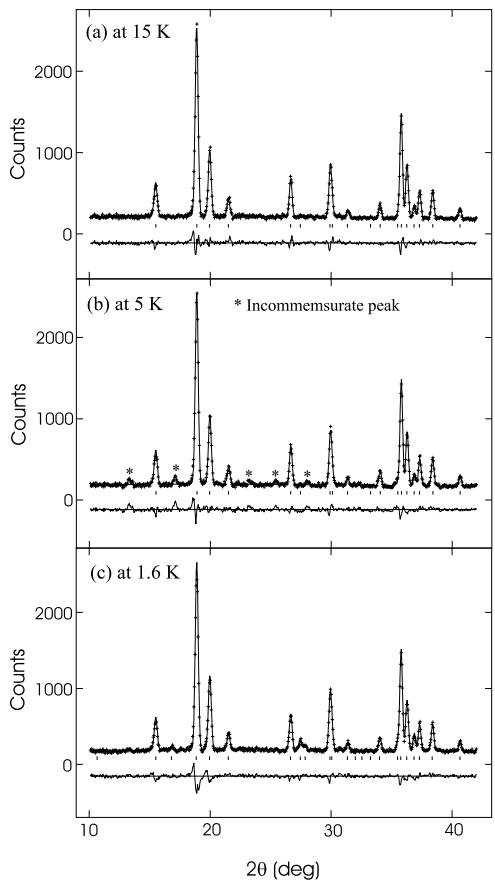}
\caption{Low-angle portions of the BT1 neutron powder diffraction
pattern collected at $15$K, $5$K, and $1.6$K.  (a) Nuclear structure
fitting. (b) Nuclear structure fitting only, incommensurate magnetic
peaks observed. (c) Both nuclear and magnetic structures were
included in the fit. The differences between observed and calculated
intensities are shown at the bottom of each figure.  The vertical
lines indicate the positions of the possible Bragg peaks.}
\label{powderspectra}
\end{center}
\end{figure}

\begin{figure}[ht]
\begin{center}
  \includegraphics[height=8cm,bbllx=85,bblly=207,bburx=500,
  bbury=620,angle=0,clip=]{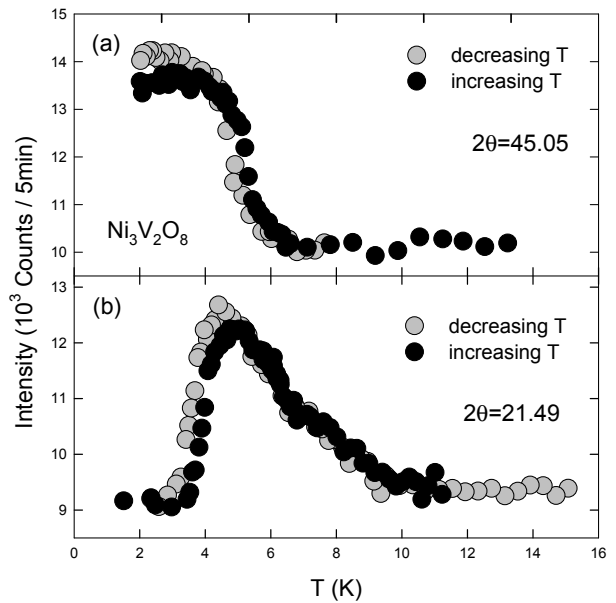}
  \caption{Top: Intensity at the scattering angle $2\Theta$,
  close to the AF peaks $(1,1,2)$ and $(1,3,0)$.
  The intensity above $4$K is related to $(0.73,3,1)$
  magnetic reflection associated with the incommensurate magnetic
  structures. Bottom: Intensity at the scattering angle $2\Theta$
  for the incommensurate magnetic peak $(0.73,1,0)$ as a function
  of temperature.}
  \label{powdertemp}
\end{center}
\end{figure}

Fig.~\ref{powderspectra} shows the low-angle neutron powder
diffraction pattern measured at $1.6$K, $5$K and $15$K. The
appearance of new Bragg peaks upon cooling indicates that the
compound undergoes transitions to magnetic order below $9.1$K. The
Bragg peaks below $4$K can be indexed with ordering wave-vectors
that are commensurate, whereas no such identification is possible
for higher temperatures, suggesting incommensurate magnetic
structures at higher temperatures. Fig.~\ref{powdertemp} shows the
temperature dependence of the intensity of an incommensurate
magnetic peak near $2\Theta= 21.49^o$, and scattering associated
with the commensurate order observed at $2\Theta=45.05^o$. This
shows that the incommensurate phase exists in a finite temperature
window between $4$ and $9.1$K, and that the low $T$ state has
commensurate magnetic order.\par

\begin{figure}[ht]
\begin{center}
  \includegraphics[height=5cm,bbllx=50,bblly=240,bburx=490,
  bbury=555,angle=0,clip=]{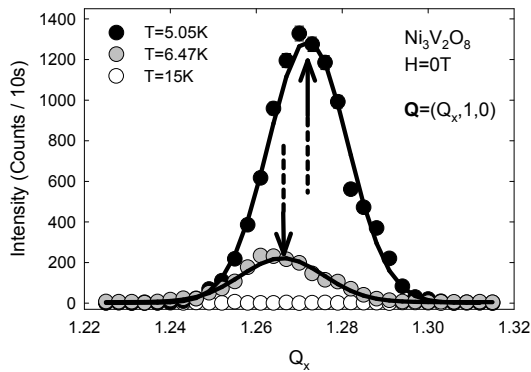}
  \caption{Neutron diffraction intensity measured as a function of $Q_x$
  for wavevector ${\bf Q}$=$(Q_x,1,0)$. Here the peak position determines
  the value of $1+q$.}
\label{incomhscan}
\end{center}
\end{figure}

The magnetic order was further investigated with experiments on a
single-crystal in which scattering was monitored only in the
$(h,k,0)$ and $(h,k,k)$ planes. Fig.~\ref{incomhscan} shows the
elastic neutron scattering at three different temperatures for
wavevectors of the form $(Q_x,1,0)$. Upon entering the magnetic
phase, a Bragg peak is formed with a maximum intensity for
$Q_x=Q_x^0 \approx 1.27$ r. l. u.'s. This result indicates weight in
the Fourier transform of the spin $S({\bf Q})$ at a wavevector
$(Q_x^0,1,0)$ which is outside the first Brillouin zone of the
primitive unit cell. We deduce that, Bragg scattering is allowed at
wavevectors
\begin{eqnarray}
{\bf G} + {\bf q} \equiv {\bf G} \pm (q,0,0) \ ,
\end{eqnarray}
where $q \approx 0.27$ and ${\bf G}$ is a reciprocal lattice vector
of the form $G=(l+m,l-m,n)$, where $l$, $m$, and $n$ are integers.
The ordering wave-vector in the HTI and LTI phases is thus ${\bf
v}=(q,0,0)$. The peak in Fig. 7 is in the $l=1,m=n=0$ zone. In this
formulation the wavevector ${\bf q}$ indicates that the spin
function varies as $\exp(i {\bf q} \cdot {\bf R})$ as the position
is displaced through a translation vector ${\bf R}$ of the lattice,
as defined in the caption to Fig. \ref{STRUCTURE}. Thus, for a given
irrep, the actual spin amplitudes are determined by the value of $q$
and the value of the spin coordinates within the unit cell as given
in Table \ref{TableIrreducibleRepInCom} (or \ref{SYMAD}).

Note that this wavevector $q$ does {\it NOT} give the phase factor
introduced upon moving from one spine site to its nearest neighbor.
In view of the intra-cell structure (given in Table
\ref{TableIrreducibleRepInCom}), different components of spin will
involve different phase factors. If we consider the ${\bf
a}$-component of spin in irrep $\Gamma_4$ and let the translation
vector ${\bf R}$ be $(X,Y,Z)$, then we see that
\begin{eqnarray}
S_{s1,a}(X)&=& e^{iq(2 \pi/a)[X+(a/4)]} m_s^a \ , \nonumber \\
S_{s4,a}(X)&=& e^{iq(2 \pi/a)[X+(3a/4)]} (-m_s^a) \ ,
\label{QZEROEQ} \end{eqnarray} where $s1$ and $s4$ are site labels
as in Fig. \ref{STRUCTURE}.  Thus $S_{s4,a}(X)/S_{s1,a}(X)= -
e^{iq \pi}$.  Similarly one finds $S_{s1,a}(X+a)/S_{s4,a}(X)= -
e^{iq \pi}$. Our conclusion is that translation along a spine by
$a/2$ introduces a phase factor $e^{i(q+1)\pi}$. So, the
wavevectors for the $a$-component along a single spine are $q_\pm
= 1 \pm q$. As we shall see later, other wave-vectors are needed
to reproduce the position dependence of other spin components
within this representation.

\begin{figure}[ht]
\begin{center}
  \includegraphics[height=7cm,bbllx=20,bblly=160,bburx=590,
  bbury=625,angle=0,clip=]{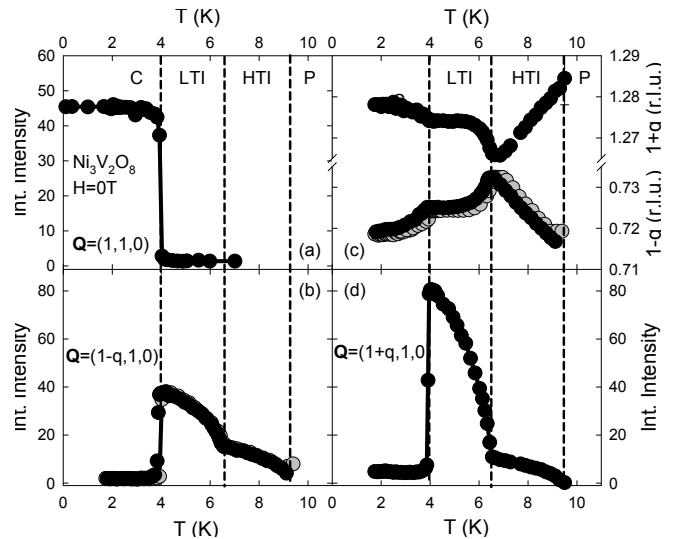}
  \caption{Temperature dependence of magnetic Bragg peaks in zero
  field. The low-temperature field was reached by zero-field
  cooling. Integrated intensities were obtained by integrating
  diffraction intensities measured as a function of $h$ wave-vector
  transfer. (a) Temperature dependence of $h$-integrated intensity
  of the AF $(1,1,0)$ reflection. (b) Temperature
  dependence of $h$-integrated intensity of the incommensurate
  $(1-q,1,0)$ reflection. (c) Temperature dependence of the
  wave-vector $(1\pm q,1,0)$ of the incommensurate magnetic reflections.
  (The data points below $T=4$K represent nonequilibrium domains
  of the LTI phase embedded in the C phase.  Such peaks are only
present after cooling through the LTI phase.)
  (d) Temperature dependence of $h$-integrated intensity of the
  incommensurate $(1+q,1,0)$ reflection.}
  \label{results0T}
\end{center}
\end{figure}

The incommensuration $q$ is weakly temperature dependent, as shown
in Fig.~\ref{results0T}c, indicating competing interactions in the
spin lattice. The temperature dependence of the integrated intensity
of the incommensurate Bragg peaks indicates the onset of magnetic
order at $T_{\rm PH}=9.1$K and further a second order transition at
about $T_{\rm HL}=6.3$K. These transition temperatures are
consistent with specific heat measurements, which show sharp peaks
at these temperatures, and with magnetization data. The existence of
two different incommensurate magnetic phases is a further indication
of competing interactions in NVO.

\begin{figure}[hb]
\begin{center}
  \includegraphics[height=5cm,bbllx=80,bblly=240,bburx=495,
  bbury=545,angle=0,clip=]{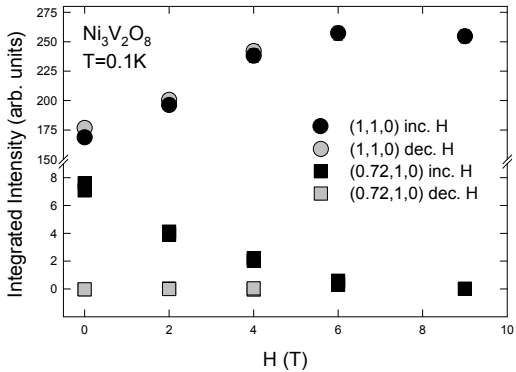}
  \caption{Integrated intensity of the $(1,1,0)$ and the $(0.72,1,0)$
  reflections for increasing and decreasing field after the sample
  was cooled in zero field.}
  \label{field0p1K}
\end{center}
\end{figure}

Fig.~\ref{results0T} shows that the incommensurate Bragg peaks
abruptly lose most of their intensity at $T_{\rm LC}=4$K, below
which temperature a commensurate magnetic order becomes dominant.
Commensurate Bragg peaks were observed at the $(h,k,0)$ for $h+k=$
even, so the commensurate structure is associated with an ordering
wave-vector ${\bf v}=(0,0,0)$. The magnetic unit cell in the C phase
is thus identical to the chemical unit cell.\par

In the zero-field cooled sample, we observed a weak incommensurate
peak which is not present in the $8$T field cooled sample. This is
evidence that the incommensurate phase below $T_{\rm LC}$ is
metastable - possibly a reflection of how close the commensurate and
incommensurate magnetic order lie in energy. However,
Fig.~\ref{field0p1K} shows that the ground state of NVO can be
annealed through the application of a field along the {\bf c}-axis
which makes the incommensurate Bragg peak vanish. The incommensurate
Bragg peaks do not reappear when lowering the field, but the
intensity is instead transferred to the commensurate peak that grows
in strength.\par

\begin{figure}[hb]
\begin{center}
  \includegraphics[height=7cm,bbllx=15,bblly=154,bburx=590,
  bbury=650,angle=0,clip=]{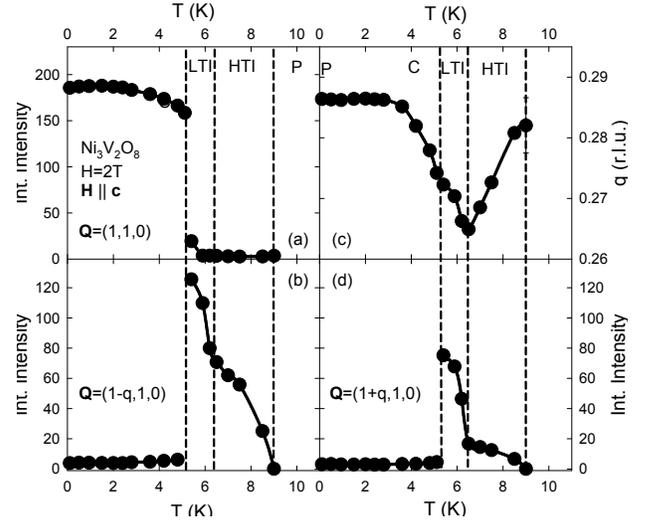}
  \caption{Temperature dependence of magnetic Bragg peaks for
  $H=2$T, applied along the crystallographic {\bf c}-axis.
  The low-temperature state was reached by cooling the sample in
  a field of $8$T. Integrated intensities were obtained
  by integrating the intensity of the magnetic Bragg reflection when
  opbserved by a scan over the $h$-component of the wave-vector transfer.
  (a) Temperature dependence of integrated intensity of the AF $(1,1,0)$
  reflection. (b) Temperature dependence of integrated
  intensity of the incommensurate $(1-q,1,0)$ reflection.
  (c) Temperature dependence of the incommensuration $q$.
  (d) Temperature dependence of integrated intensity of the
  incommensurate $(1+q,1,0)$ reflection.}
  \label{results2T}
\end{center}
\end{figure}

\subsection{Field dependence of Bragg reflections}
The field dependence of the magnetic Bragg reflections was
investigated only with fields along the {\bf c}-axis.
Fig.~\ref{results2T} shows the magnetic reflections at $2$T as a
function of temperature. Upon heating, the $(1,1,0)$ reflection
disappears in a first-order transition as the incommensurate $(1 \pm
q,1,0)$ reflections appear. The commensurate phase survives to
higher temperatures than at zero field, and the ordered moment
increases with field, both indications that the magnetic field
stabilizes the commensurate order. The LTI magnetic structure
occupies a relatively narrow temperature range while the temperature
boundaries for the HTI magnetic structure are nearly independent of
field.

\begin{figure}[hb]
\begin{center}
  \includegraphics[height=7cm,bbllx=9,bblly=154,bburx=590,
  bbury=650,angle=0,clip=]{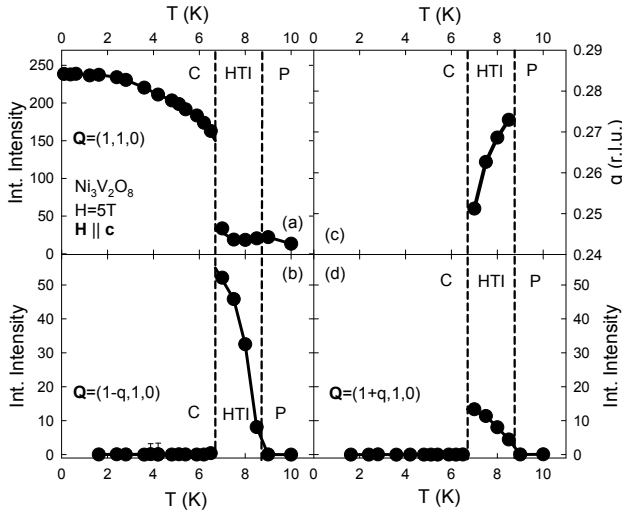}
  \caption{As for Fig. \protect{\ref{results2T}}, but
for $H=5$T.}
  \label{results5T}
\end{center}
\end{figure}

The LTI magnetic structure is further suppressed with increasing
magnetic field along the {\bf c}-axis. At $5$T, as shown in
Fig.~\ref{results5T}, the LTI structure does not occur, and as the
temperature is increased, the commensurate structure gives way
directly to the HTI magnetic structure. The phase transition between
the paramagnetic and HTI phase occurs at a temperature somewhat
below its zero-field critical temperature $T_{\rm PH}=9$K. It is for
this field direction that the phase boundaries depend most strongly
on the field. The magnetic phase boundaries obtained with these
neutron measurements are consistent with the phase diagram obtained
through specific heat measurement with the field along the {\bf
c}-axis.\par

\begin{figure}[hb]
\begin{center}
  \includegraphics[height=7cm,bbllx=15,bblly=154,bburx=590,
  bbury=650,angle=0,clip=]{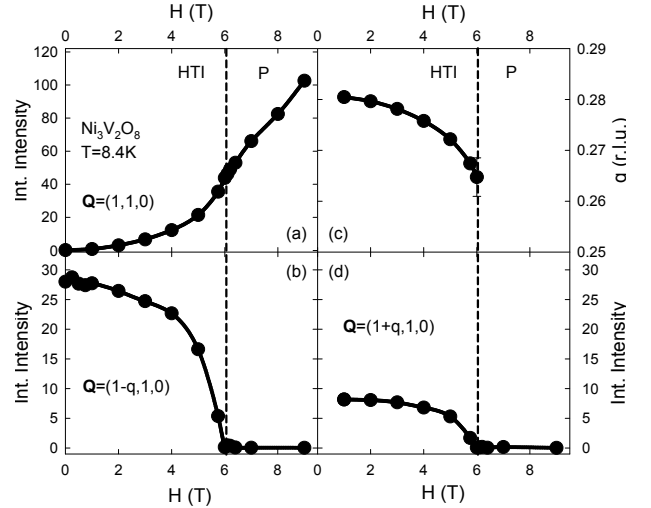}
  \caption{Field dependence of magnetic Bragg peaks at $T=8.4$K.
  Integrated intensities were obtained as for
  Fig. \protect{\ref{results2T}}. Field dependence of
  the integrated intensity of (a) the AF $(1,1,0)$ reflection,
  (b) the incommensurate $(1-q,1,0)$ reflection and (d)
  the incommensurate $(1+q,1,0)$ reflection. (c) shows the
  field dependence of the incommensurate wave vector $q$.}
  \label{field8p4T}
\end{center}
\end{figure}

A field along the ${\bf c}$ axis in the HTI phase leads to
suppression of the incommensurate Bragg peak at $(1-q,1,0)$ and an
increase in the intensity of the $(1,1,0)$ reflection, as shown in
Fig. \ref{field8p4T} for $T=8.4$K. The temperature dependence of the
intensity of the incommensurate Bragg peaks suggests that the HTI
phase disappears in a continuous phase transition at a critical
field $H_c$, giving way to a commensurate field driven AF phase at
higher fields. This contrasts with the first order nature of the
phase boundary between the LTI and AF phases. As shown in Fig.
\ref{field8p4T}, both the incommensurate wave vector, $q$, and the
integrated intensity of the (1,1,0) AF Bragg peak increase
progressively more rapidly as the field increases.

\subsection{Phase diagram} The zero-field phase boundaries at
$T_{\rm PH}$, $T_{\rm HL}$ and $T_{\rm LC}$ deduced from the
diffraction experiments are consistent with those observed with
specific heat measurements. In contrast, the intensity of the
$(1,1,0)$ does not show any anomaly at $T_{\rm CC'}$, to a level of
$0.5\%$. This suggests that the CC' phase transition leaves the
magnetic structure of C-phase described by ordering wave-vector
${\bf v}=(0,0,0)$ unaltered.\par

Figures \ref{results2T} and \ref{results5T} already show that the
C-LTI  and the C-HTI transitions are first order at small $H$.
Similar results were obtained when we varied $H$ at fixed $T$.
Specifically, Fig. \ref{stagmom} shows the jumps in the staggered
moment of the C phase as one moves from the LTI or from the HTI
phases into the C phase, for $T<8$K. In contrast, at higher
temperatures and fields the transition from the HTI to the P phase
is continuous, as can be seen from Fig. \ref{field8p4T}. In fact, at
finite field along ${\bf c}$ one cannot really distinguish between
the P and the C phases, and we already saw in Figs. \ref{results2T}
and \ref{results5T} that the HTI-P transition is continuous for
$H\leq 5$T. We thus conclude that the HTI-C transition changes from
being continuous to being first order somewhere around the top of
the boundary of the HTI phase in Fig. \ref{PHASED}. Such {\it
tricritical} points are abundant in anisotropic antiferromagnets
subject to magnetic fields.\par

\begin{figure}[ht]
\begin{center}
  \includegraphics[height=5cm,bbllx=12,bblly=185,bburx=595,
  bbury=565,angle=0,clip=]{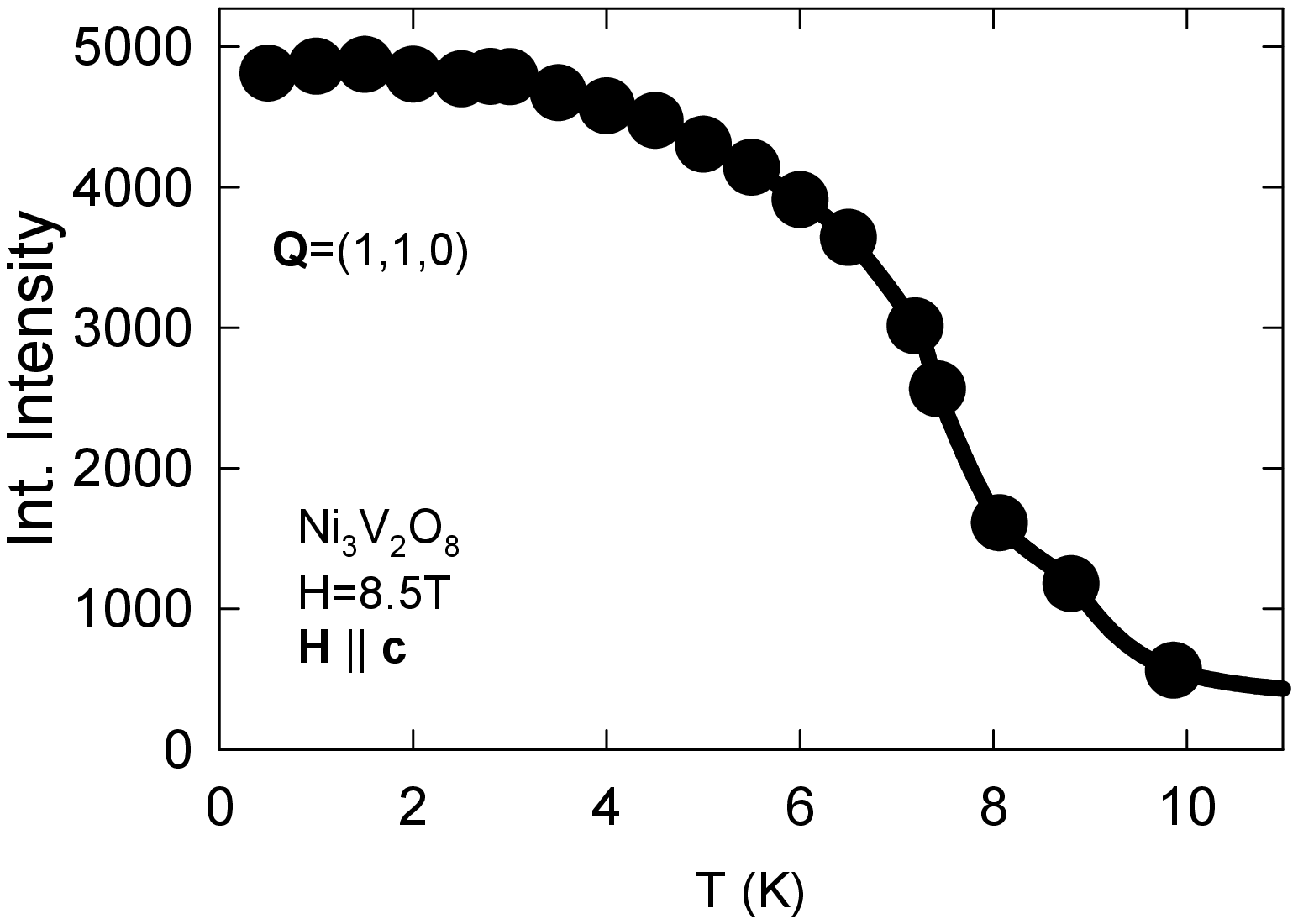}
  \caption{Temperature dependence of $h$-integrated intensity
  of the AF $(1,1,0)$ reflection in a field
  of $H=8.5$T, applied along the {\bf c}-axis, showing
  the absence of a second order phase transition when cooling from the
  paramagnetic phase to the low-temperature commensurate phase.
  The data instead reflect a cross over phenomenon for $T \approx 8$K.}
  \label{results8p5T}
\end{center}
\end{figure}

The change in slope of the $(110)$ intensity versus $T$ curve at the
HTI-P transition (see Fig.~\ref{field8p4T}) indicates the coupling
between the commensurate and incommensurate order parameters, which
will be discussed in Sec.~\ref{SubSectionPhaseBoundaries}.\par

The observations which are relevant for the C phase are (a) the
specific heat data at 9T  which show no anomaly and (b) the lack of
any detectable anomaly in the $T$-dependence of the(110) peak
intensity (see Fig.~\ref{results8p5T}). As will be shown in
Sec.~\ref{structuresymmetry}, the P phase in a field allows the
development of order characteristic of only the irrep $\Gamma_7$ in
which there is a uniform moment along ${\bf c}$ and a staggered
moment along {\bf a}. This suggests that the magnetic structure in
the C and P phase (in a field) exhibit the same symmetry, and that
the symmetry at high field at low and high temperature is the same
as that of the C phase at zero field because it can be reached
without crossing a second order phase transition.\par

\subsection{Magnetic Structures}
Even more detailed information about the spin interactions can be
obtained by determining the symmetry of the ordered magnetic
structures. In this paper, we will focus our attention on the HTI,
LTI, C and P phases, and we will leave a discussion of the C' phase
for a later publication.

\subsubsection{High-temperature incommensurate (HTI) structure}
For temperatures between $T_{\rm HL}$ and $T_{\rm PH}$, Bragg
reflections occur at the $(2n+1 \pm q,2m+1,0)$ and $(2n+1\pm
q,2m+1,2m+1)$ positions. The intensities of $170$ magnetic Bragg
reflections were measured and can be explained with a magnetic
structure belonging to representation  $\Gamma_4$ given in
Table~\ref{SYMAD}. The magnetic structure at $T=7$K is given by
\begin{eqnarray}
\tilde {\bf m}_s^{\rm \Gamma_4} &=&(1.9(1),0.2(1),0.2(2))\;\mu_B\nonumber\\
\tilde {\bf m}_c^{\rm \Gamma_4} &=&(0,0.0(1),0.2(2))\;\mu_B \ ,
\label{HTISTR} \end{eqnarray} where the number in parenthesis is the
uncertainty in the last digit quoted. The quality of the fit is
given by $R_{\rm p}$=$0.15$ and $\chi^2$=$12$ (see
Table~\ref{Table1} for the definition of these quantities). The spin
arrangement at $T=7$K is illustrated in Fig. \ref{MAGSTR}b. It is
predominantly a modulated structure on the spine sites with moments
that are parallel to the {\bf a} direction. The moments on the
cross-tie sites either vanish or are very small. [If the cross-tie
moments in the {\bf b} direction are nonzero, they are out of phase
with the spine moments because of the phase factor introduced by Eq.
(\ref{III})].

\begin{figure}[ht]
\begin{center}
  \includegraphics[height=5cm,bbllx=80,bblly=240,bburx=495,
  bbury=545,angle=0,clip=]{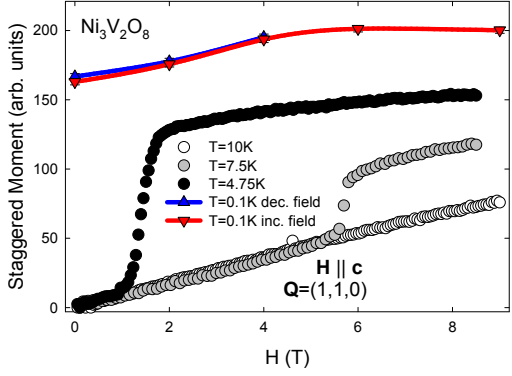}
  %{stagmoment.eps}
  \caption{Field dependence of effective staggered moment at
  three different temperatures, obtained from the $(1,1,0)$
  reflection by taking the square root of the peak intensity.}
  \label{stagmom}
\end{center}
\vspace{0.6 in}
\end{figure}

To determine the HTI magnetic structure in a magnetic field along
the {\bf c}-axis, we measured the intensity of $28$ magnetic Bragg
peaks in the $(h,k,0)$ plane for $T$=$8$K and $H$=$5$T. The data is
best described by the basis vectors of the irrep $\Gamma_4$, and the
magnetic structure is given by
\begin{eqnarray}
\tilde m_s^{\Gamma_4} &=& (1.60(4), 0.08(3),0(2))\mu_B\nonumber \\
\tilde m_c^{\Gamma_4} &=& (0,0.02(5),0(2))\mu_B
\end{eqnarray}The quality of the fit is given by $R_{\rm p}$=$0.14$
and $\chi^2$=$6.7$. The effect of the field along the {\bf c}-axis
is thus to reduce the ordered moment along the {\bf a}-axis compared
to the zero-field structure at a somewhat lower temperature. This
may be because the field induces a uniform moment along the {\bf
c}-axis which reduces the amount of moment available along the {\bf
a}-axis.\par

\subsubsection{Low-temperature incommensurate (LTI) structure}
\begin{figure}[hb]
\includegraphics[height=11cm,bbllx=70,bblly=295,bburx=475,
  bbury=810,angle=0,clip=]{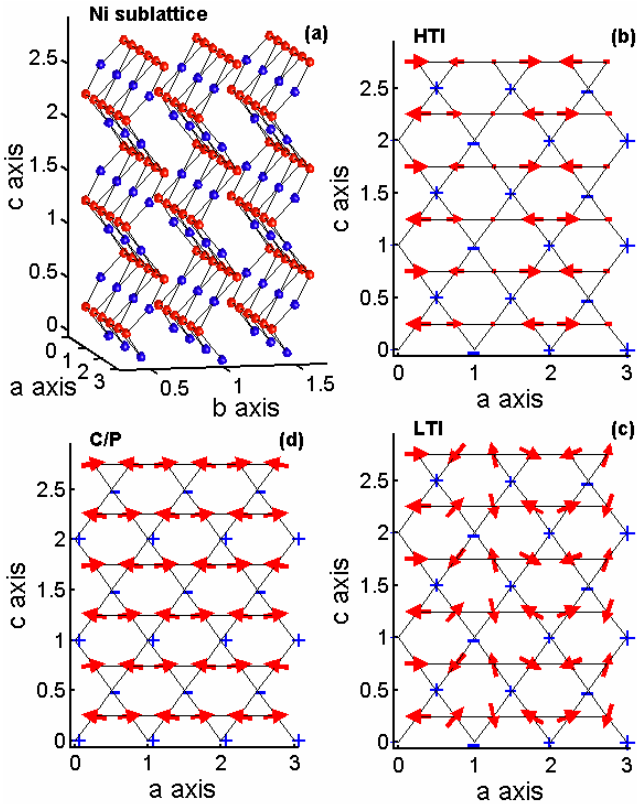}
\caption{\label{MAGSTR}Lattice (a) and magnetic structure (b-d) of
various phases. The size of the $+$ and $-$ signs correspond to the
components out of the page and into the page, respectively. In panel
(d) the canting of the C phase is magnified for legibility. For the
HTI and LTI phases, the ${\bf a}$-component has a wavelength
approximately $1.4a$ so that  $q \approx 0.27$, as indicated in Fig.
\protect{\ref{incomhscan}}.}
\end{figure}

For temperatures between $T_{\rm LC}$ and $T_{\rm HL}$, Bragg
reflections were observed at the $(2n+1 \pm q,2m+1,0)$ and $(2n+1\pm
q,2m+1,2m+1)$ positions. These are the same Bragg peaks as observed
in the HTI phase, but their relative intensity has changed,
indicating that the spins undergo a spin reorientation at $T_{\rm
HL}$. The present diffraction data are consistent with either a
$\Gamma_1 + \Gamma_4$ or a $\Gamma_2+\Gamma_4$ structure.  We choose
the former spin structure, not because it has a smaller value of
$\chi^2$, but because it is consistent with the appearance of
ferroelectricity, as is discussed in Sec.~\ref{ferro}, below.  With
that assumption the magnetic structure at $5$K is given by
\begin{eqnarray}
\tilde {\bf m}_s^{\rm \Gamma_4}&=&(1.6(1),0.03(10),0.01(7))\;\mu_B\nonumber\\
\tilde {\bf m}_s^{\rm \Gamma_1}&=&(0.0(5),1.3(1),0.1(1))\;\mu_B\nonumber\\
\tilde {\bf m}_c^{\rm \Gamma_4}&=&(0,1.4(1),-0.04(9))\;\mu_B\nonumber\\
\tilde {\bf m}_c^{\rm \Gamma_1}&=&(-2.2(1),0,0)\;\mu_B\, .
\label{LTISTR} \end{eqnarray} The quality of the fit is given by
$R_{\rm p}$=$0.19$ and $\chi^2$=$7$. The corresponding spin
structure is shown in Fig. \ref{MAGSTR}c and consists of elliptical
${\bf a}-{\bf b}$ plane spirals on spine and cross-tie sites.\par

The structure at $T=5$K thus consists of spirals on the spine and
cross-tie sites, propagating along the {\bf a}-axis with moments in
the {\bf a}-{\bf b}-plane, as shown in Fig. \ref{MAGSTR}d. An
inspection of the spin structure suggests that nn interactions
between Ni on adjacent spines are AF, both within and between
Kagom\'{e} planes. The moment in the {\bf c}-direction is zero
within the error bar, indicating the presence of a spin anisotropy
which forces the spin into the {\bf a}{\bf b} plane.\par

To determine the LTI magnetic structure in a magnetic field along
the {\bf c}-axis, we measured the intensity of $28$ magnetic Bragg
peaks in the $(h,k,0)$ plane for $T$=$6$K and $H$=$2$T. We obtained
best agreement with the experimental data for basis vectors
belonging to the irreps $\Gamma_1$ and $\Gamma_4$. The structure is
given by
\begin{eqnarray}
\tilde {\bf m}_s^{\rm \Gamma_4}&=&(2.5(1),-0.1(2),0(2))\;\mu_B\nonumber\\
\tilde {\bf m}_s^{\rm \Gamma_1}&=&(-0.5(3),1.1(1),0(2))\;\mu_B\nonumber\\
\tilde {\bf m}_c^{\rm \Gamma_4}&=&(0,0.0(1),0(2))\;\mu_B\nonumber\\
\tilde {\bf m}_c^{\rm \Gamma_1}&=&(0.06(8),0,0)\;\mu_B\, ,
\end{eqnarray}The quality of the fit is given by
$R_{\rm p}$=$0.29$ and $\chi^2$=$9.3$. Qualitatively, these
parameters are similar to those of the zero-field structure. We
remind the reader that Eq. (\ref{III}) implies a phase difference of
$\pi/2$ between the ${\bf a}$- and ${\bf b}$-components on the spine
sites. The high-field LTI phase thus consists of a spiral on the
spine sites and no moment on the cross-ties, possibly because a
transverse field has a strong effect on the cross-ties which are
more weakly coupled antiferromagnetically. In contrast to the HTI
phase, however, the ordered moment at non-zero field is higher
moment than at zero field.

\subsubsection{P and C phase}
The C phase has the same symmetry as the P phase, and for fields
along the {\bf c}-axis, there is no phase boundary between the
high-field phase and the zero-field phase for $T_{\rm CC'}<T<T_{\rm
LC}$. The symmetry of the C magnetic structure can thus be
determined in a high magnetic field along the {\bf c}-axis. We
measured a set of magnetic Bragg peaks in the $(h,k,0)$ plane at
$T=0.1$K and $H=8$T, and we found that the data is best described by
irrep $\Gamma_7$ and the parameters
\begin{eqnarray}
    {\bf m}_s^{\rm \Gamma_7} & =& (2.4(1),0,0.0(5))\;\mu_B\nonumber\\
    {\bf m}_c^{\rm \Gamma_7} & =& (0,0.8(1),0(1))\;\mu_B\, .
\label{H0STR}
\end{eqnarray}The quality of the fit is given by $R_{\rm p}$=$0.23$
and $\chi^2$=$17$. When we allowed the fit to include also
$\Gamma_1$ parameters, the components of the magnetization were
found to be $0.1(1) \mu_B$ and therefore statistically
insignificant.\par

A magnetic field along the {\bf c}-axis induces magnetic order even
in the paramagnetic phase. At $T=10$K and $H=8$T, the spin structure
is best described by the irrep $\Gamma_7$ with the following
parameters:
\begin{eqnarray}
    {\bf m}_s^{\rm \Gamma_7} & =& (0.61(2),0,0.0(2))\;\mu_B\nonumber\\
    {\bf m}_c^{\rm \Gamma_7} & =& (0,0.37(3),0.0(4))\;\mu_B\, .
\label{G7STRUCTURE}
\end{eqnarray}The quality of the fit is given by $R_{\rm
p}$=$0.31$ and $\chi^2$=$2.49$. Thus the field-induced magnetic
structure at $T=10$K is described by the same irrep as at $T=0.1$K
and $H=8$T. This is corroborated by the absence of a phase
transition upon cooling in a high field, as shown in the temperature
dependence of the $(1,1,0)$ Bragg peak shown in
Fig.~\ref{results8p5T}.\par

\subsection{Ferroelectricity}\label{ferro}
Recently NVO has been shown to have remarkable ferroelectric
behavior.\cite{FERRO} A spontaneous polarization ${\bf P}$ has been
found to appear only in the LTI phase. In other words, upon cooling
into the LTI phase, the ferroelectric order parameter develops
simultaneously with the LTI magnetic order parameter. In Ref.
\onlinecite{FERRO} it was proposed that the multi-component order
parameter associated with this phase transition requires the
trilinear coupling, $V$, where
\begin{eqnarray}
V &=& \sum_\gamma \Biggl[ a_\gamma \sigma_{\rm LTI}^* \sigma_{\rm
HTI} + a_\gamma^* \sigma_{\rm LTI} \sigma_{\rm HTI}^* \Biggr]
P_\gamma \ . \label{TRI} \end{eqnarray} Here $P_\gamma$ is the
$\gamma$-component of the spontaneous polarization and the
$\sigma$'s are complex-valued order parameters which describe the
incommensurate long range order associated with irrep $\Gamma_4$ for
the HTI phase and with irrep $\Gamma_1$ for the additional order
parameter appearing in the LTI phase. Of course, $V$ must be
invariant under the symmetry operations of the paramagnetic
phase.\cite{IED,LL}  As we have seen, the magnetic order parameters
satisfy Eq. (\ref{SYMEQ}), which here is
\begin{eqnarray}
{\cal I} \sigma_{\rm A} &=& \sigma_{\rm A}^* \ ,
\end{eqnarray}where A denotes either LTI or HTI. Using this
relation, we see that the invariance of $V$ under spatial inversion
implies that $a_\gamma$ is pure imaginary: $a_\gamma = i r_\gamma$,
where $r_\gamma$ is real valued. Thus we may write
\begin{eqnarray}
V &=& 2 \sum_\gamma r_\gamma P_\gamma |\sigma_{\rm LTI}
 \sigma_{\rm HTI}| \sin(\phi_{\rm HTI}-\phi_{\rm LTI}) \ ,
\end{eqnarray}
where the phases are defined by $\sigma_{\rm A} = |\sigma_{\rm A}|
e^{i \phi_{\rm A}}$.\par

To be invariant under the operations of the little group $P_\gamma$
must transform like $\Gamma_4 \otimes \Gamma_1$. Referring to
Table~\ref{incomcharactertable}, we see that this means that
$P_\gamma$ has to be odd under $2_a$ and $m_{ac}$. The first of
these conditions means that $r_\gamma$ can only be nonzero for
$\gamma=b$ or $\gamma=c$. The second condition means that $r_\gamma$
can only be nonzero for $\gamma=b$, in agreement with the
observation\cite{FERRO} that the spontaneous polarization only
appears along the {\bf b}-direction. Had we chosen the irrep
$\Gamma_2$ for the new LTI representation in Eq. (\ref{LTISTR}), we
would have incorrectly predicted the spontaneous polarization to be
along the ${\bf c}$-direction.\par

\subsection{Magnetization and Susceptibility Measurements}
\begin{figure}
\includegraphics[height=8.5cm]{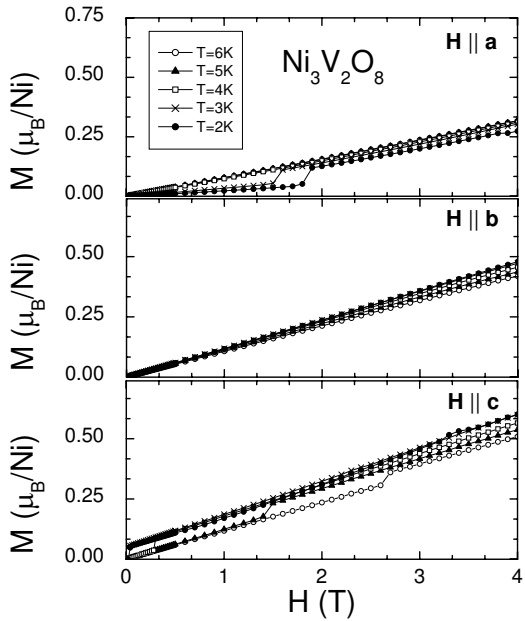}
\caption{$M$ versus $H$ along the three crystallographic axes
for a sequence of temperatures.}
\label{MvH}
\vspace{0.3 in}
\end{figure}

\begin{figure}
\includegraphics[height=7cm]{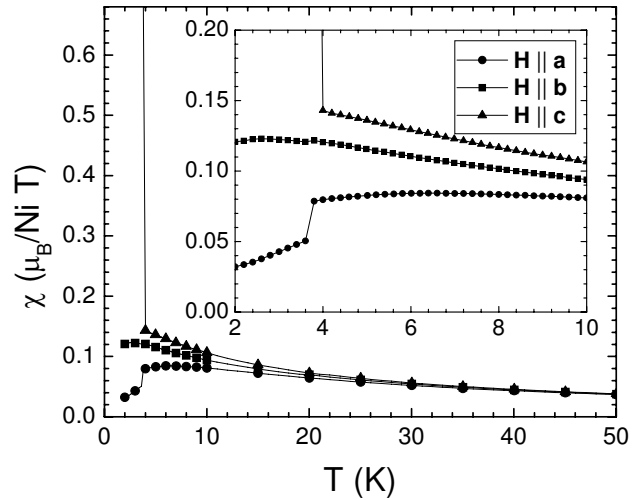}
\caption{$\chi$ for magnetic field applied along each of the
crystallographic directions in Bohr magnetons per Ni ion per
Tesla versus $T$, where $\chi$ is defined by $\chi = M(H=0.1$T$)/H$.}
\label{chi}
\end{figure}

Although neutron diffraction enables one to fix many details of the
magnetic structure, it is hard to obtain the bulk magnetization from
these measurements because the signal associated with bulk
magnetization is buried in the nuclear Bragg peaks. Accordingly, we
summarize here the results for the zero wavevector magnetization,
$M$, measured with a SQUID magnetometer, as a function of field for
fields along each of the crystallographic directions shown in Figs.
\ref{MvH} and \ref{chi}.\par

The following features are noteworthy. In Fig.~\ref{MvH} one sees
that for $H$ along the {\bf c}-axis, there is a range of temperature
(corresponding to the C phases) in which $M$ vs $H$ does not
extrapolate to zero for $H \rightarrow 0$. This is the best
measurement of the weak FM moment in this phase. The C' phase may
also have a finite remnant magnetization though further measurements
are needed there. From these data one sees confirmation of the phase
boundaries obtained from specific heat measurements (shown in Fig.
\ref{PHASED}) which here are signaled by discontinuities in the
magnetization as a function of $H$ (when a phase boundary is
crossed.) In Fig. \ref{chi} we show $M/H$, measured for a small
field $H=0.1$T. This quantity will be nearly equal to the zero field
susceptibility except when it probes the spontaneous magnetization,
as when ${\bf H}$ is along (0,0,1) and $T<4$K and the system is in
the C or C' phase. It is remarkable that there are no visible
anomalies associated with the phase transitions involving the HTI
phase.\par

\section{Theoretical Interpretations}
Here we discuss the simplest or ``minimal" model that can explain
the experimental results for NVO. In general, the higher the
temperature the simpler the model needed to explain experimental
results. Accordingly, we discuss the theoretical models for the
phases in the order of decreasing temperature. That way, we will see
that the minimal model is constructed by sequentially including more
terms as lower temperature phases are considered.

\subsection{The high temperature incommensurate (HTI) phase}

\subsubsection{Competing nearest and next nearest neighbor interactions}
Incommensurate phases usually result from competing interactions. In
the HTI phase we found spin ordering predominantly on the spine
sites [see Eq. (\ref{HTISTR})], so the minimal model will deal only
with the spine spins. Since the incommensurate wave vector is along
the spine axis, it is reasonable to infer that this
incommensurability arises from competition between nn and nnn
interactions along a spine. We already indicated that such
competition is plausible, in view of the nearly 90$^{\rm o}$ angle
of the nn Ni-O-Ni bonds. Thus the minimal model to describe the
incommensurate phase is
\begin{eqnarray}
{\cal H}_{\rm spine} &=& {1 \over 2} \sum_{n=1}^2\sum_\alpha
J_{n\alpha} \sum_{{\bf r} , \delta_n} S_\alpha({\bf r})
S_\alpha({\bf r} + \delta_n) \nonumber\\
&+& {\cal H}_{\rm A} + {\cal H}_{ss'} \ , \label{HTIEQ}
\end{eqnarray} where $J_{1\alpha}$ and $J_{2\alpha}$ represent the
({\it a priori} anisotropic) nn and nnn exchange interactions,
$\alpha$ is summed over the axes $a,~b$ and $c$, ${\bf r}$ is
summed over only spine sites, and $\delta_n= \pm (n/2)a \hat {\bf
a}$ are the first and second neighbor displacement vectors for the
spine sites (remember that there are two spine spins along each
spine axis in the unit cell, so the nn distance is $a/2$).
%Except for specific points,
%it suffices to consider the isotropic exchange limit, where we
%shall use $J_{n\alpha} \equiv J_n$.
Also ${\cal H}_{ss'}$ represents (probably weak) AF interactions
between nn in adjacent spines: $J_{b\alpha}$ for spins at a distance
$b/2$ in the {\bf b}-direction and $J_{c\alpha}$ for spins at a
distance $c/2$ in the {\bf c}-direction. (In this minimal model, for
the purpose of Fourier transformation, we place the lattice sites on
an orthorhombic lattice for which the ${\bf a}-{\bf c}$ planes are
{\it not} buckled, but we retain the same interactions between spins
as for the buckled lattice. In this way our model gives the correct
thermodynamics, even though it should not be used to obtain
scattering intensities.)\par

To have competition, we must have $J_{2\alpha}>0$ (at least for the
relevant $\alpha$, which turns out to be $\alpha=a$). Given that at
low temperature we end up with a commensurate AF ordering of the
{\bf a}-spin components, it is also reasonable to assume that
$J_{1\alpha}>0$ (at least for $\alpha=a$). If the spines did not
interact with one another, they would form an array of independent
one-dimensional systems for which thermal and/or quantum
fluctuations would destroy long-range order. To understand the
coupling between adjacent spines, note that a displacement $({\bf
b}/2)$ (from a spine in one Kagom\'{e} layer to a spine in an
adjacent Kagom\'{e} layer) takes site \#1 in one unit cell into site
\#4 in another unit cell and vice versa. Similarly this displacement
takes site \#2 in one unit cell into site \#3 in another unit cell
and vice versa. From Table VII (or VIII) one see that for the active
irreps \#4 and \#1 such a nearest neighbor displacement along ${\bf
b}$ corresponds to a change in the signs of $m_{sa}$ and $m_{sb}$.
The fact that these are the dominant order parameters for the spine
sites therefore strongly suggests that the nn interspine
interactions along ${\bf b}$ are antiferromagnetic. The evidence for
antiferromagnetic interactions along ${\bf c}$ is almost as
compelling. For these nn pairs (sites \#1 and \#2, or sites \#3 and
\#4) the largest order parameters of the LTI phase (see Eq.
(\ref{LTISTR}))B, i. e. the $x$ components of spine spins ordered
according to irrep \#4 and the $y$ component ordered according to
irrep \#1 are both antiferromagnetic along ${\bf c}$. While it is
true that the ordering of the $x$-component according to irrep \#1
is ferromagnetic along ${\bf c}$, this component of the order
parameter is indistinguishable from zero. Thus we conclude that the
nn interactions $J_{1\alpha}$ for all directions $\alpha$ are
antiferromagnetic.\par

Also, in Eq. (\ref{HTIEQ}) ${\cal H}_{\rm A}$ is a single ion
anisotropy,
\begin{eqnarray}
{\cal H}_A=-\sum_\alpha A_\alpha\sum_{\bf r}S_\alpha({\bf r})^2\ .
\label{HA}
\end{eqnarray}The continuum mean-field phase diagram of this
Hamiltonian is known, for both exchange and single ion
anisotropies.\cite{Nagamiya,Kaplan} To determine which phase orders
as the temperature is lowered from the paramagnetic phase, it is
sufficient to look at the quadratic terms in the expansion of the
free energy per spin in the Fourier components of the spins,
\begin{eqnarray}
F^{\rm spine}_2={1 \over 2} \sum_{\bf p}\sum_\alpha
[\chi_{s,\alpha} ({\bf p})]^{-1} S_\alpha({\bf p})
S_\alpha(-{\bf p})\ ,
\label{F2}
\end{eqnarray}
where
\begin{eqnarray}
[\chi_{s,\alpha}({\bf p})]^{-1}= T/C +{\hat J}_\alpha({\bf p})-2A_\alpha
\label{chialph}
\end{eqnarray}is the $\alpha$-component of the inverse susceptibility
of the spine sites associated with the Fourier component
\begin{eqnarray}
S_\alpha({\bf p})=\sum_{\bf r}S_\alpha({\bf r})
e^{-i{\bf p}\cdot {\bf r}}/{\cal N}
\end{eqnarray}(with the sum over {\it all} the ${\cal N}=4N_{uc}$ spine
spins in the lattice), while ${\hat J}_\alpha({\bf p})$ is the
Fourier transform of the exchange interactions. In
Eq.~(\ref{chialph}), $C$ is the Curie constant,
\begin{eqnarray}
C=S(S+1)/3=2/3.\label{Curie}
\end{eqnarray}
(We measure energy in temperature units, which amounts to setting
the Boltzmann constant $k=1$.) For our simple model
(\ref{HTIEQ}),
\begin{eqnarray}
{\hat J}_\alpha({\bf p}) &=& 2[J_{1\alpha}\cos(p_aa/2)
+J_{2\alpha}\cos(p_aa)\nonumber\\
&+&J_{b\alpha}\cos(p_bb/2)+J_{c\alpha}\cos(p_cc/2)]\ .
\label{Jalph}
\end{eqnarray}As $T$ is lowered, the first phase to order will
involve the order parameter $S_\alpha({\bf p})$ for which
$[\chi_{s,\alpha} ({\bf p})]^{-1}$ first vanishes. For
$J_{1\alpha}>4|J_{2\alpha}|$, this happens for
\begin{eqnarray}
{\bf p}={\bf Q}_0 \equiv 2\pi(1/a,1/b,1/c),\label{QQQ}
\end{eqnarray}implying a simple two sublattice antiferromagnet
along each spine chain (with the orthorhombic unit cell containing
four spins in each sublattice, and with the spins inside the unit
cell varying with $e^{i{\bf Q}_0 \cdot{\bf r}}$). However, for
$J_{2\alpha}>|J_{1\alpha}|/4$, the minimum in ${\hat J}_\alpha({\bf
p})$ occurs at the incommensurate wave vector
\begin{eqnarray}
{\bf p}_{0\alpha}=2\pi(q_{0\alpha}/a,1/b,1/c).
\label{PZERO}
\end{eqnarray}The modulation wave vector $q_{0\alpha}$ for the
$\alpha$ spin component (in r.l.u.'s) is given by
\begin{eqnarray}
\cos (\pi q_{0\alpha}) &=& - J_{1\alpha}/(4J_{2\alpha}) \ ,
\label{QEQN}
\end{eqnarray}and thus the susceptibility for this wavevector becomes
\begin{eqnarray}
[\chi_{s,\alpha} ({\bf p}_{0\alpha})]^{-1}&=& T/C-
2[J_{2\alpha}+J_{1\alpha}^2/(8J_{2\alpha})\nonumber\\
&+&J_{b\alpha}+J_{c\alpha}+A_\alpha].
\end{eqnarray}Experimentally, we know that the leading order parameter
in the HTI phase concerns $S_a({\bf p}_{0a})$.  In this phase, the
neutron diffraction data also give a wave vector which varies
slightly with temperature, close to $q_{0a}= 1 \pm q \approx 0.72$
or 1.28. To avoid confusion, from now on we use the notation $q_0
=0.72$.  Using this approximate value, Eq.~ (\ref{QEQN}) gives
\begin{eqnarray}
J_{1a} \approx 2.55  J_{2a} .\label{J1J2}
\end{eqnarray}

Thus we conclude that at this mean field level one has
\begin{eqnarray}
T_{\rm PH}= 2C[J_{2a}+J_{1a}^2/(8J_{2a})+J_{ba}+J_{ca}+A_a],
\label{TPHEQ}
\end{eqnarray}
and we end up with a longitudinally modulated spin structure,
with $S_b({\bf r}) = S_c({\bf r})=0$ and, along a single spine,
\begin{eqnarray}
S_a(x) = S \cos(2 \pi q_0x/a+\phi_n),
\end{eqnarray}where the phase $\phi_n$ may depend on the spine index
$n$. The $n$-dependence of $\phi_n$ would be determined by the
inter-spine coupling. For our model the neutron diffraction data
indicates that adjacent spines are antiferromagnetically arranged,
so that for a general spine site we may write
\begin{eqnarray}
S_a({\bf r})=S \cos({\bf p}_0 \cdot {\bf r}+ \phi_0)\ , \label{SX}
\end{eqnarray}where the transverse components of ${\bf p}_0$ are
fixed as in Eq. (\ref{PZERO}) and we adopt the notation ${\bf
p}_0\equiv {\bf p}_{0a}$. When one goes beyond continuum mean-field
theory, it is found that instead of $q_0$ being a continuous
function of $J_{1a}$ and $J_{2a}$ one obtains a `devil's staircase'
dependence of the wave vector on the control
parameters.\cite{Aubry,ANNNI} This treatment also shows that
$\phi_0$ can not be fixed arbitrarily (as it would be in continuum
mean-field theory). Furthermore, critical fluctuations will reduce
the actual $T_{\rm PH}$ by a factor which depends on the spatial
anisotropy.

\subsubsection{Dzyaloshinskii-Moriya interactions}\label{secDM}

In addition to the leading {\bf a}-component of the spine spins, Eq.
(\ref{HTISTR}) also indicates a small, but non-negligible, spine
moments along the {\bf b}-axis and along the {\bf c}-axis. Such
moments follow directly from the Dzyaloshinskii-Moriya (DM)
interactions\cite{Dzyaloshinskii,Moriya} between spine spins. For
simplicity, we consider here only the nn DM interactions along the
spine,
\begin{eqnarray}
{\cal H}_{DM}=\sum_{\bf r}{\bf D}({\bf r})\cdot {\bf S}({\bf
r})\times {\bf S}({\bf r}+a{\hat {\bf a}}/2),
\end{eqnarray}and the symmetry of the lattice dictates that the
DM vectors behave as
\begin{eqnarray}
{\bf D}({\bf r})=\bigl (0,D_be^{i{\bf Q}_0\cdot {\bf r}},D_c
e^{i{\bf P}_c\cdot {\bf r}}\bigr ),
\end{eqnarray}

\noindent where ${\bf Q}_0$ is the wave vector for the
two-sublattice commensurate wave vector [Eq. (\ref{QQQ})], while
${\bf P}_c=(0,0,2\pi/c)$ represents AF ordering along the {\bf
c}-axis. Next nearest neighbor DM interactions, discussed in II,
do not change the qualitative results presented here. The Fourier
representation of ${\cal H}_{DM}$ yields a free energy of the form
\begin{eqnarray}
&&F_{DM}=\sum_{\bf p}
\bigl [D_b\cos(p_a a/2)S_a({\bf p})S_c(-{\bf p}-{\bf Q}_0)\nonumber\\
%&-&S_a({\bf p})S_c(-{\bf p}-{\bf Q}_0)]\nonumber\\
&&-iD_c\sin(p_a a/2)[S_a({\bf p})S_b(-{\bf p}-{\bf P}_c)]+h.c..
%\nonumber\\
%&-&S_b({\bf p})S_a(-{\bf p}-{\bf P}_c)].
\label{FDM}
\end{eqnarray}

Introducing quadratic terms in the transverse spine spin
components, as in Eq. (\ref{F2}), we can now minimize the free
energy and find these transverse components: a non-zero $S_a({\bf
p}_0)$ generates
\begin{eqnarray}
S_c({\bf p}_0+{\bf Q}_0)&\approx&- 2D_b\chi_{s,c}({\bf p}_0+{\bf Q}_0)
\cos(\pi q_0)S_a({\bf p}_0),
\nonumber \\
S_b({\bf p}_0+{\bf P}_c)&\approx& 2iD_c \chi_{s,b}({\bf p}_0+{\bf P}_c)
\sin(\pi q_0)S_a({\bf p}_0),
\label{SbSc}
\end{eqnarray}
consistent with the signs and phases of $\tilde m_s^b$ in Table
\ref{SYMAD}. In fact, for all the group representations, all the
internal structure within the orthorhombic unit cell can be
reproduced using factors like $e^{i{\bf p}\cdot{\bf r}}$, with the
three wave vectors ${\bf p}_0$, ${\bf p}_0+{\bf Q}_0$ and ${\bf
p}_0+{\bf P}_c$ (with possibly different ${\bf p}_0$'s for
different representations).

Using the values from Eq. (\ref{HTISTR}), and $p_a = 2\pi q_0/a$
with $q_0 \approx 0.72$, Eq. (\ref{SbSc}) yields
\begin{eqnarray}
D_c &\approx &-0.07 [\chi_{s,b} ({\bf p}_0+{\bf P}_c)]^{-1},\nonumber\\
D_b &\approx &0.04 [\chi_{s,c} ({\bf p}_0+{\bf Q}_0)]^{-1}.\label{DbDc}
\end{eqnarray}
    %Our mean field expressions for the susceptibilities imply that
    %$[\chi_a({\bf p}_0)]^{-1}=k(T-T_{\rm PH})$, while (for isotropic
    %exchange, $J_{n\alpha}\equiv J_n$) $[\chi_b({\bf p}_0+{\bf
    %P}_c)]^{-1}-[\chi_a({\bf p}_0)]^{-1}=4J_c-2A_b+2A_a$ and
    %$[\chi_c({\bf p}_0+{\bf Q}_0)]^{-1}-[\chi_a({\bf
    %p}_0)]^{-1}=-4J_1\cos(\pi q_0)+4(J_b+J_c)-2A_c+2A_a$. Since
    %$T_{\rm PH} \approx 9$K, it is reasonable to assume that the order
    %of magnitude of all the $J$'s and therefore also of all the
    %inverse non-ordering susceptibilities is also $\sim 10$K. This
    %gives $D_b \sim -0.4$K and $D_c\sim -0.6$K.
Near $T_{\rm PH}$ , $[\chi_{s,a} ({\bf p}_0)]^{-1}$ is small, but
$[\chi_{s,b} ({\bf p}_0+{\bf P}_c)]^{-1}$ and $[\chi_{s,c}({\bf p}_0
+{\bf Q}_0)]^{-1}$ remain finite. Since $T_{\rm PH} = 9.1$K, it is
reasonable to assume that all these inverse susceptibilities are of
order 10K. Thus we have the estimates, $D_b \sim 0.4$K and $D_c \sim
-0.7$K. Below we conjecture a set of exchange parameters, which give
$[\chi_b({\bf p}_0+{\bf P}_c)]^{-1} \sim 13$K and $[\chi_c({\bf
p}_0+{\bf Q}_0)]^{-1} \sim 47$K, changing these estimates into $D_b
\sim 0.5$K and $D_c \sim -3$K.

\subsubsection{The spine-cross-tie interactions}\label{secPD}

The representation $\Gamma_4$ also allows some small incommensurate
moments along the {\bf b} and {\bf c} axes on the cross-tie sites.
Indeed, the experimental values in Eq. (\ref{HTISTR}) also allow for
such moments, alas with large error bars. Here we discuss the
possible theoretical origin for these moments, namely the
anisotropic spin interactions between the spine and the cross-tie
spins. In all the phases, the two spine chains which are nearest
neighbors to a given row of cross-tie sites in an ${\bf a}-{\bf c}$
plane (in this section we ignore the buckling, which does not affect
the present considerations) have antiparallel {\bf a}-components of
spins. Thus even in the incommensurate phases, the isotropic nn
spine-cross-tie interaction is frustrated, and one needs to add
(symmetric and antisymmetric) {\it anisotropic} spine-cross-tie
interactions. In the simplest approach, these interactions can be
written in terms of an effective internal field produced by the four
spine spins ($s1$ to $s4$) surrounding a cross-tie spin $c2$ (see
Fig. \ref{STRUCTURE}), so that
\begin{eqnarray}
{\cal H}_{sc}=-\sum_\alpha H^c_\alpha S^c_\alpha(c2),\label{Hxtie}
\end{eqnarray}
with
\begin{eqnarray}
H^c_\alpha=\sum_\beta \sum_{s1}^{s4}{\cal
J}_{\alpha\beta}(i)S_\beta(i), \label{CALJEQ}
\end{eqnarray}
where the matrix ${\cal J}(i)$ (which relates to the coupling
between the spins at $si$ and at $c2$) contains both symmetric
(pseudo-dipolar, PD) and antisymmetric (DM) off-diagonal terms,
whose signs depend on $i$ (see II). Ignoring the interactions
among cross-tie spins, each such spin will follow its local field,
\begin{eqnarray}
S_\alpha^c(c2)=\chi_{c,\alpha} H^c_\alpha,\label{xspin}
\end{eqnarray}
where $\chi_{c,\alpha}$ is the $\alpha$-component of the cross-tie
susceptibility.

The above analysis yields the spin components on each cross-tie
site in terms of the four surrounding spine spins. Taking also
into account the variation of the matrices ${\cal J}(i)$ for
different plaquettes, and assuming only linear response for the
cross-tie spins, we end up with
\begin{eqnarray}
S_c^c({\bf p}_0)&=&-4\chi_{c,c}(j_{ac}-d_b)S_a({\bf p}_0)\sin(\pi q_0/2),\nonumber\\
S_b^c({\bf p}_0+{\bf P}_c)&=&4\chi_{c,b}(j_{ab}+d_c)S_a({\bf
p}_0)\sin(\pi q_0/2)\ ,
\label{xmoment}
\end{eqnarray}
where $j_{\alpha\beta}$ and $d_\gamma$
are the symmetric (PD) and anti-symmetric (DM) elements of the
matrix ${\cal J}(1)$. Using the experimental values of the
cross-tie moments from Eq. (\ref{HTISTR}), we end up with
\begin{eqnarray}
(j_{ab}+d_c)\chi_{c,b} &=& - 0.06 \pm 0.06 \ , \nonumber \\
(j_{ac}-d_b)\chi_{c,c} &=&   0 \pm 0.06 \ .\label{jd}
\end{eqnarray}
Ignoring the very weak interaction among the cross-tie spins, we
can use the free spin Curie susceptibility, $\chi_{c,\alpha}
\approx 1/T \sim 0.2/$K, hence $(j_{ab}+d_c)\approx -0.3(3)$K,
$(j_{ac}-d_b) \approx 0.0(3)$. However, the uncertainties of these
estimates are very large.

\subsection{The low temperature incommensurate (LTI) phase}

Mean-field theory\cite{Nagamiya,Kaplan} also indicates that, if the
uniaxial anisotropy energy is not too large, then as the temperature
is reduced further below $T_{\rm PH}$, a second continuous phase
transition occurs at $T_{\rm HL}$, below which transverse modulated
order appears, leading to an incommensurate state with elliptical
polarization. Again, this seems to be consistent with the
experiments: Eq. (\ref{LTISTR}) indicates a growing $b$-component on
the spine spins and a growing $a$-component on the cross-tie spins.
For simplicity, we again concentrate only on the spine spins. One
can understand this transition intuitively by realizing that as the
temperature is lowered the fixed length constraint on spins
(embodied in the quartic terms in the Landau expansion) becomes
progressively more important. Since the experiment indicates
transverse ordering in the {\bf b}-direction [see Eq.
(\ref{LTISTR})], we conclude that the anisotropies still require
that $S_c({\bf r})=0$, and we write a Landau expansion in both
$S_a({\bf p})$ and $S_b({\bf p})$. A priori, the quadratic terms in
$|S_a({\bf p})|^2$ and in $|S_b({\bf p})|^2$ could have minima at
slightly different wave vectors ${\bf p}_1$ and ${\bf p}_2$
respectively [see Eq. (\ref{QEQN})]. The quartic coupling between
the two spin components would then be of the form $|S_a({\bf
p}_1)|^2|S_b({\bf p}_2)|^2$. Depending on the ratio of the amplitude
of this term to those of $|S_a({\bf p}_1)|^4$ and $|S_b({\bf
p}_2)|^4$, one could end up with a second transition, at $T_{\rm
HL}<T_{\rm PH}$, at which $S_b({\bf p}_2)$ would begin to order
(this is similar to the usual appearance of a tetracritical point,
see e.g. Ref. \onlinecite{TCP}). However, the quartic terms also
contain a term of the form $[S_a({\bf p}_1)^2S_b(-{\bf
p}_2)^2+h.c.]\delta({\bf p}_1-{\bf p}_2) \equiv 2|S_a({\bf
p}_1)|^2|S_b({\bf p}_2)|^2 \cos[2(\phi_a-\phi_b)]\delta({\bf
p}_1-{\bf p}_2)$, where we have written $S_\alpha({\bf
p})=|S_\alpha({\bf p})|e^{i\phi_\alpha}$. Usually, the coefficient
in front of this term is positive, and therefore it is minimized
when $\cos[2(\phi_a-\phi_b)]=-1$, and it lowers the free energy only
if ${\bf p}_1={\bf p}_2$. If the values of ${\bf p}_1$ and ${\bf
p}_2$ are sufficiently close to one another, then the presence of
this term results in locking the wave vectors to each other: ${\bf
p}_1={\bf p}_2={\bf p}_0$, consistent with the experiment.\par

Thus, within this theory we expect spin ordering of the type
\begin{eqnarray}
S_a({\bf r})={\tilde S}_a \cos({\bf p}_0 \cdot {\bf r}+ \phi_a),\nonumber\\
S_b({\bf r})={\tilde S}_b \cos({\bf p}_0 \cdot {\bf r}+ \phi_b)\,
, \label{Sperp}
\end{eqnarray}
with ${\tilde S}_b$ growing continuously from zero as the
temperature is lowered through $T_{\rm HL}$, and with
$2(\phi_a-\phi_b)$ an odd integer multiple of $\pi$. The fact that
the two order parameters are out of phase can be understood
intuitively: if longitudinal and transverse spin components are
combined, they are closer to obeying the fixed length constraint
if they are out of phase with one another.

Within the Landau theory, we would expect $|{\tilde S}_a|^2$ to
grow at $(T_{\rm PH}-T)$ down to $T_{\rm HL}$. Below that
temperature, $|{\tilde S}_b|^2$ grows linearly with $(T_{\rm
HL}-T)$, while the slope of $|{\tilde S}_a|^2$ versus $T$
decreases. \cite{TCP} Qualitatively, this is what one observes in
Figs. \ref{results0T}, \ref{results2T} and \ref{results5T}.

\begin{figure}[ht]
\begin{center}
\includegraphics[height=8cm,bbllx=20,bblly=30,bburx=530,
  bbury=530,angle=0,clip=]{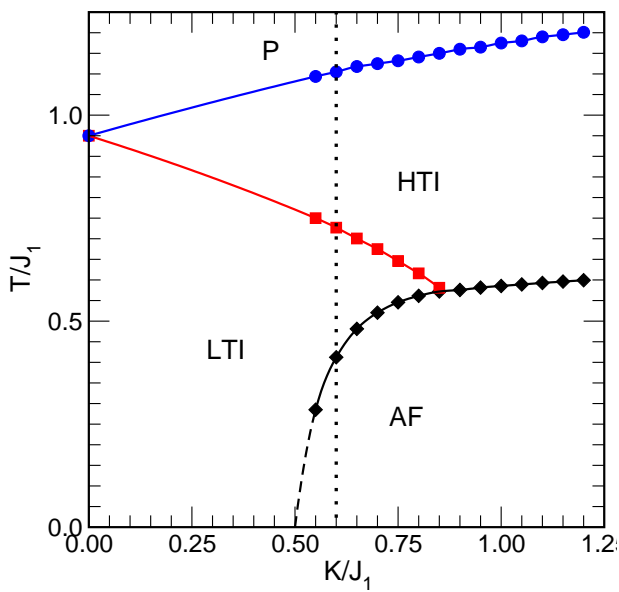}
\caption{Mean-field phase diagram for the isotropic $J_1$-$J_2$
model with easy-axis anisotropy scaled by $K\equiv A_a$. Here ``HTI"
denotes a longitudinally polarized incommensurate phase and ``LTI"
an elliptically polarized incommensurate phase.  In both phases the
modulation vector is given by Eq. (\protect{\ref{QEQN}}). ``AF"
denotes a two-sublattice collinear AF phase. For large anisotropy
this model reduces to the ANNNI model.\protect{\cite{ANNNI}}}
\label{PHASETZ}
\end{center}
\end{figure}

The above theory ignores higher harmonics of the incommensurate
order parameters. Although these harmonics may be negligible close
to $T_{\rm PH}$, they may grow at lower temperatures. Therefore,
we went beyond the Landau expansion and numerically minimized the
mean field moments on long chains of spin-1 ions with the
interaction (\ref{HTIEQ}), with an isotropic exchange which has
$J_1=2.55 J_2$ and with a uniaxial single-ion energy $A_a=K$ (see
II for details). The resulting phase diagram, shown in Fig.
\ref{PHASETZ}, is qualitatively consistent with the above picture.
It is intuitively clear that the range in temperature over which
the LTI phase is stable decreases as the anisotropy is increased.
(Large anisotropy disfavors the existence of transverse spin
components.)

Having derived the leading order parameter, we can now evaluate
secondary spin components, arising due to either the spine-spine DM
interaction or the spine-cross-tie PD interactions. Basically, such
an analysis will generate all the other spin components that are
allowed by representations $\Gamma_4$ and $\Gamma_1$. In the above
discussion, we represented $\Gamma_4$ by its largest component
$S_a({\bf p}_0)=m_{sa}^{\Gamma_4}$, and $\Gamma_1$ by $S_b({\bf
p}_0)=m_{sb}^{\Gamma_1}$. Once this is done, one could in principle
deduce more information on the PD and DM coupling constants.
Unfortunately, the only data available in the LTI phase is at 5K,
and these data show  apparently large values of the cross-tie spins.
These large values are surely beyond our linear response treatment,
and therefore we are not able to use them for identifying the
coupling constants. At the moment, we do not understand this
apparent fast saturation of the cross-tie spins.

\subsection{The commensurate antiferromagnetic C phase}

\subsubsection{Structure and symmetry}\label{structuresymmetry}

As noted in Refs. \onlinecite{Nagamiya,Kaplan}, and as confirmed in
our mean field analysis (Fig. \ref{PHASETZ}), lowering the
temperature yields a first order transition between the LTI phase
and a commensurate phase. In this commensurate phase, the spine
spins return to be along the ${\bf a}$-axis (like in the HTI phase),
but now they all have the same length. At zero field we identify
this phase with the C phase. At finite field along the ${\bf c}$,
this phase coincides with the P phase. For our minimal model ${\cal
H}_{\rm spine}$, the order parameter in the P phase is the simple
two sublattice staggered moment, $N_{s,a} \equiv S_a({\bf
Q}_0)$.\par

In addition to gaining energy from the fixed length constraint, one
also gains energy from the appearance of a weak FM moment. This
moment arises due to the anisotropic PD and/or DM interactions.  We
have already noted that the symmetry of the P phase is confirmed by
the lack of a phase boundary to the paramagnetic phase at high
magnetic field along the {\bf c} direction. (In a magnetic field
along {\bf c}, the paramagnetic phase must be invariant under a
two-fold rotation about the {\bf c}-axis and under inversion because
${\bf H}$ is an axial vector. Thus the magnetic structure must be
invariant under these operations but should change sign under
two-fold rotations about the {\bf a} or {\bf b}-axis. This means
that in the paramagnetic phase with a field applied along the {\bf
c} axis we must have only the representation $\Gamma_7$ of Table
\ref{TableIrreducibleRepCom}.) The above conclusions are supported
by the structure observed at $T=0.1$K and $H=8$T.\par

A similar symmetry analysis shows that at high temperature we have
only the irrep $\Gamma_5$ of Table \ref{TableIrreducibleRepCom} for
${\bf H}$ along {\bf a} and $\Gamma_3$ for ${\bf H}$ along {\bf b}.
Thus, for ${\bf H}$ along {\bf a} or {\bf b}, a comensurate phase
containing $\Gamma_7$ can only be reached by crossing a phase
boundary. This is also consistent with the data: for fields along
the {\bf a}- or {\bf b}-axis, the paramagnetic phase and the C phase
are always separated either by the HTI or by both the HTI and LTI
phases.\par

As stated, an external uniform field $H$ along {\bf c} generates a
nonzero magnetization on both the spine and cross-tie sites. The DM
and PD interactions then generate a nonzero staggered moment on the
spine sites, $N_{s,a}$. In the paramagnetic phase, this moment will
be approximately linear in the field. This explains the nonzero
intensity at ${\bf Q}=(110)$ in the P phase, seen in Figs.
\ref{results5T}, \ref{results8p5T} and \ref{field8p4T}. Within the
Landau theory, symmetry allows a biquadratic coupling of this
staggered moment to the incommensurate order parameters, e.g.
$N_{s,a}^2|S_a({\bf p}_0)|^2$. Near the P-HTI transition, this term
renormalizes $[\chi_a({\bf p}_0)]^{-1}$ by an amount of order $H^2$,
thus shifting $T_{\rm PH}$ by such an amount.\par

At higher $H$, the coupling of the incommensurate HTI order
parameter to both $M$ and $N$ also renormalizes the quartic term
$|S_a({\bf p}_0)|^4$, yielding a tricritical point when this term
turns negative.\par

Below $T_{\rm PH}$, $|S_a({\bf p}_0)|^2$ grows as $(T_{\rm PH}-T)$.
The above quartic term then induces a corresponding change in the
inverse staggered susceptibility $[\chi_a({\bf Q}_0)]^{-1}$, causing
the change in slope of $N_{s,a}$ versus $T$ seen in Fig.
\ref{results5T}. More of these calculations are given in II.

\subsubsection{Weak ferromagnetic moment}
\label{WEAK}

Now we discuss the possible sources of the small FM moment observed
in the C phase and possibly also the C' phase. This can arise either
from the DM interactions\cite{Dzyaloshinskii,Moriya} or from the
pseudo-dipolar
interaction.\cite{ChouAharony,YildirimHarris,SachidanandamYildirim,kastner}
We start by considering the DM interaction. Equation (\ref{SbSc})
(with $q_0$ replaced by 1) shows that when ${\bf p}_0$ is replaced
by ${\bf Q}_0$ then the spine staggered moment $N_{s,a}$ generates a
FM moment
\begin{eqnarray}
M_{s,c} \approx 2D_b\chi_{s,c}(0)N_{s,a}.
\end{eqnarray}
(From now on we distinguish between spine and cross-tie properties
by the subscripts $s$ and $c$).

Using our rough estimates $D_b\sim 0.4$K and $\chi_{s,c} \sim 0.1$K,
and ignoring the spine-cross-tie coupling (see below), this gives
$M_{s,c}\sim 0.08N_{s,a} \sim 0.12 \mu_B$. This spine value is
within the error bar given for this moment in Eq.~(\ref{H0STR}). The
same equation also allows for a similar moment on the cross-tie
sites. That same equation then also implies a FM on the cross-ties
of order $M_{c,c}\approx -0.1\mu_B$.

Similarly, the off-diagonal exchange spine-cross-tie interaction
will generate a FM moment in the {\bf c}-direction on the cross-tie
sites: Eq. (\ref{xmoment}) implies \begin{eqnarray}
M_{c,c}=-4\chi_{c,c}(j_{ac}-d_b)N_{s,a}, \end{eqnarray} and the
estimates $(j_{ac}-d_b) =0\pm 0.3$K and $\chi_{s,c} \sim 0.1$K give
$M_{c,c}=0\pm 0.2\mu_B$, which are consistent with the above
estimate.

These estimates are justified only if we ignore the direct exchange
coupling between the FM moments ${\bf M}_s$ and ${\bf M}_c$.
Although this coupling cancelled out for the AF and incommensurate
moments, the diagonal elements in the matrices ${\cal J}(i)$ do add
up for the FM moments. Assuming isotropic diagonal elements, this
generates a coupling $4J_{\rm av}{\bf M}_s \cdot {\bf M}_c$ per
cross-tie site, where $J_{\rm av}$ is the spine-cross tie nn
interaction. A complete analysis then requires a minimization of the
free energy with respect to the two FM moments. The situation here
is very similar to that which occurred in Sr$_2$Cu$_3$O$_4$Cl$_2$
\cite{kastner} which also has a weak FM moment in an AFM phase and
involves two types of magnetic ions (copper). Following Ref.
\onlinecite{kastner}, we assume that at low field in the C phase the
staggered moment on the spine sites is practically saturated. The
free energy of the small FM moments per Ni ion can then be written
as
\begin{eqnarray}
3F=\sum_\alpha
\Bigl[\frac{M_{s,\alpha}^2}{\chi_{s,\alpha}}+\frac{M_{c,\alpha}^2}{2\chi_{c,\alpha}}\Bigr]
 -2{\bf H}_s \cdot {\bf M}_s-{\bf H}_c \cdot {\bf
M}_c,
\end{eqnarray}where we denote $J_{\rm pd}=j_{ac}-d_b$ and
\begin{eqnarray}
{\bf H}_s&=&{\bf H}-2D_bN_{s,a}{\hat c},\nonumber\\
{\bf H}_c&=&{\bf H}-4 J_{\rm av}{\bf M}_s-4J_{\rm pd}N_{s,a}{\hat
c}.
\end{eqnarray}Minimization with respect to the magnetizations then
yields the total magnetization per Ni ion,
\begin{eqnarray}
M_\alpha=M_0 \delta_{\alpha,c}+\chi_\alpha H_\alpha, \label{MvsH}
\end{eqnarray}with the average zero-field moment
\begin{eqnarray}
M_0&=&\frac{4}{3}N_{s,a}\bigl[\chi_{s,c}\frac{1-2\chi_{c,c}J_{\rm
av}}{1-8J_{\rm av}^2\chi_{s,c}\chi_{c,c}}\nonumber\\
&&\times (4J_{\rm av}J_{\rm pd}\chi_{c,c}-D_b)-J_{\rm
pd}\chi_{c,c}\Bigr ]
\end{eqnarray}
and susceptibility
\begin{eqnarray}
\chi_\alpha=\frac{2\chi_{s,\alpha}+\chi_{c,\alpha}-8J_{\rm
av}\chi_{s,\alpha}\chi_{c,\alpha}}{3(1-8J_{\rm
av}^2\chi_{s,\alpha}\chi_{c,\alpha})}.\label{chial}
\end{eqnarray}
Note that For $J_{\rm av}=0$, this reduces to the trivial
$M_\alpha=(2/3)M_{s,\alpha}+(1/3)M_{c,\alpha}$.\par

Experimentally, we seem to observe a weak temperature dependence of
$M_0$, with a stronger $T$-dependence in $\chi_{s,\alpha}$. In the C
phase, the spine spins are ordered antiferromagnetically, and
therefore the $\chi_{s,\alpha}$'s have a weak $T$-dependence (both
for the transverse and the longitudinal susceptibilities).
Therefore, the temperature dependence in the above expressions
probably comes mainly from $\chi_{c,\alpha}$. Expanding $M_0$ to
leading order in $\chi_{c,c}$, we note that this order vanishes if
$J_{\rm pd}=2\chi_{s,c}J_{\rm av}D_b$. In the AF C phase,
$\chi_{s,c}$ is the transverse susceptibility, which is of order
$1/[4(J_1-J_2)+2(J_b+J_c+A_a)]$. It is also reasonable to assume
that $J_{\rm av}$ is of order $J_1$, given the similar geometry of
the nn ss and sc bonds. This would imply that $J_{\rm av}\chi_{s,c}$
is of order one. Since we also found that $D_b$ and $J_{\rm pd}$ are
of the same order of magnitude,  the above relation may in fact hold
approximately, and this could explain the weak $T$-dependence of
$M_0$. However, this is all highly speculative at this stage.

\subsection{Phase boundaries in the $H$-$T$ plane}
\label{SubSectionPhaseBoundaries}

\subsubsection{Qualitative discussion}

Now that we have identified the magnetic structure of all the
ordered phases, we can give a qualitative discussion of the various
boundaries in the phase diagram. In the various incommensurate
phases, the free energy must be an even function of the magnetic
field. Since the field generates a uniform magnetic moment, it
competes with the incommensurate order. Therefore, we expect the
lines between the high temperature paramagnetic phase and the HTI
phase to behave at low field like $T_{\rm PH}(H) \approx T_0- AH^2$,
with the coefficient $A$ depending on details. As stated above, this
coefficient arises due to the biquadratic coupling of the HTI order
parameter to $N_{s,a}^2$ and to $M^2$. The data seem consistent with
this, showing a small value of $A$. The observed phase boundaries
between HTI and LTI also seem almost field independent, remaining
roughly at a constant distance below $T_{\rm PH}(H)$. To leading
order, the instability which yields the transverse incommensurate
order is mostly dominated by the size of the longitudinal order
parameter, which depends mainly on this temperature difference.

We next discuss the (first order) boundaries between the phase C and
the two incommensurate phases. We start with the two upper panels in
Fig.~\ref{PHASED}, where the field is not along the weak FM moment
in phase C. A guiding principle in this discussion is that the
Zeeman energy of an antiferromagnet (or a general incommensurate
ordered state) in a uniform field, $-(1/2) \chi H^2$, is larger in
magnitude when the field is perpendicular to the staggered moment
than when it is parallel to that moment (because $\chi_\perp >
\chi_\parallel$). Now consider the $H$-dependence of the transition
temperature $T_{\rm LC}$ from the LTI phase to the C phase. When the
field is along (100), the commensurate phase gains very little
Zeeman energy (due to $\chi_\parallel$), whereas for the
incommensurate phase, the field is partly transverse, because the
spins are in a cone. Indeed, Fig.~\ref{PHASED}a shows a transition
from C to LTI as $H$ increases, with a phase boundary which might be
parabolic. In contrast, when the field is along (010), the
commensurate phase gains the entire energy $-(1/2)\chi_\perp H^2$,
whereas in the incommensurate phase, to the extent that the spins
are partially in the {\bf b}-direction, the lowering of energy is
less than that in the commensurate phase. We thus expect a
transition from the LTI phase to the C phase as $H$ along (010) is
increased, consistent with Fig.~\ref{PHASED}b. Again, the phase
boundary seems parabolic.\par

In contrast, when the field is along (001), the commensurate phase
is strongly favored because in addition to the energy involving the
transverse susceptibility, the external field also couples linearly
to the spontaneous FM moment, which is along the {\bf c}-direction.
This explains why the C-LTI and the C-HTI phase boundaries are
practically linear in the $H$-$T$ plane only for ${\bf H}$ along
(001).
%Indeed, this argument reinforces our interpretation that any possible
%spontaneous moment in the {\bf b}-direction is small.

As discussed above, the smearing of the transition between the
paramagnetic phase and the C phase results from a bilinear coupling
between the magnetization $M_c$ and the AF order parameter
$N_{s,a}$, implying that a uniform field along (001) also acts as a
staggered field on $N_{s,a}$. Had it not been for this coupling,
then the competition between HTI and C would result in a bicritical
phase diagram, with the two second order lines (P-HTI and P-C)
meeting the first order HTI-C line at a bicritical point. Indeed,
Fig.~\ref{PHASED}c does show the typical inward curvatures of the
former two lines, typical for such phase diagrams.\cite{TCP}

\subsubsection{ Field along (001)}

We now make these arguments more quantitative. We start with the
LTI-C and HTI-C phase boundaries. Since these represent first order
transitions, it suffices to compare the free energies in the
respective phases. Since the susceptibilities may have different
values in the different phases, we use the superscripts $C$ and $I$
for the commensurate and the incommensurate phases. We start with
the field along (001).

In the commensurate C phase, with a field along (001), we write
the energy per Ni ion as
\begin{eqnarray}
&&E_C=E_C^0-\frac{1}{3}{\tilde \chi}_{s,c}^C(\mu_c
H-H_s)^2-\frac{1}{6}\chi_{c,c}(H-H_c)^2,\nonumber\\
&&=E_{C}^0-\frac{1}{2}\chi_c^C H^2-H M_0-\frac{1}{3}{\tilde
\chi}_{s,c}^CH_s^2-\frac{1}{6}\chi_{c,c}H_c^2, \label{EC}
\end{eqnarray} where $\mu_\alpha=1-2J_{\rm av}\chi_{c,\alpha}$,
${\tilde \chi}_{s,\alpha}^C=\chi_{s,\alpha}^C/(1-8J_{\rm
av}^2\chi_{s,\alpha}^C\chi_{c,\alpha})$, $H_s=(2D_b-8J_{\rm
av}J_{\rm pd}\chi_{c,c})N_{s,a}$, $H_c=-4J_{\rm pd}N_{s,a}$,
$M_0=(2\mu_c {\tilde \chi}_{s,c}^CH_s+\chi_{c,c}H_c)/3$. Also, the
total susceptibility is $\chi_z^C=\frac{1}{3}(2\mu^2{\tilde
\chi}_{s,c}^C+\chi_{c,c})$. It is easy to check that these
expressions agree with those in Sec.~\ref{WEAK}. At low
temperatures, our isotropic $J_1-J_2-A$ model yields
$E_C^0=\frac{2}{3}(-J_1+J_2-A)$, where $A\equiv A_a$ is the
coefficient of the single ion anisotropy, Eq. (\ref{HA}), and
$\chi_{s,c}^C=\chi_{s,\perp}^C$ is temperature independent.

In the incommensurate phase LTI, we may have different values for
$\chi_{s,\alpha}$, which we now denote $\chi_{s,\alpha}^{I}$. As
we don't expect $\chi_{c,\alpha}$ to change, we now have the total
susceptibility as
\begin{eqnarray}
\chi_\alpha^I&=&(2\mu_\alpha^2{\tilde
\chi}_{s,\alpha}^I+\chi_{c,\alpha})/3\nonumber\\
&=& \frac{2\chi_{s,\alpha}^I+\chi_{c,\alpha}-8J_{\rm
av}\chi_{s,\alpha}^I\chi_{c,\alpha}}{3(1-8J_{\rm
av}^2\chi_{s,\alpha}^I\chi_{c,\alpha})}.
\end{eqnarray}
%with the notations  ${\tilde
%\chi}_{s,\alpha}^I=\chi_{s,\alpha}^I/(1-8J_{\rm
%av}^2\chi_{s,\alpha}^I\chi_{c,\alpha})$ and $\mu_\alpha=1-2J_{\rm
%av}\chi_{c,\alpha}$.
Since in this phase, $N_{s,a}$ is replaced by an oscillating term,
the ``spontaneous" moment on the spine spins, $M_0$, will be
replaced by an oscillating term which will give no average
contribution to the energy. For the same reason, we need to
replace $H_s^2$ and $H_c^2$ by their averages over these
oscillations, which we denote by $\gamma H_s^2$ and $\delta
H_c^2$. Thus we write
\begin{eqnarray}
E_I=E_I^0-\frac{1}{2}\chi_c^I H^2-\frac{1}{3}\gamma{\tilde
\chi}_{s,c}^IH_s^2-\frac{1}{6}\delta\chi_{c,c}H_c^2.
\end{eqnarray}
 For our $J_1-J_2-A$ model, at
low temperatures, we have
$E_I^0=\frac{2}{3}[-J_2-J_1^2/(8J_2^2)-A/2]$.

The first order transition between these two phases will occur
when $E_C=E_I$. Experimentally, it turns out that the difference
$\chi_c^C-\chi_c^I$ is very small.  Neglecting this difference,
remembering that the experimental $M_0$ seems practically
temperature independent, and neglecting any temperature dependence
of $\chi_{s,c}^C$ and $\chi_{s,c}^I$, we obtain the
transition field $H_{C,I}$ at temperature $T$ as
\begin{eqnarray}
M_0H_{C,I}&=&\Bigl({\tilde \chi}_{s,c}^C(T_1)-{\tilde
\chi}_{s,c}^C(T)\nonumber\\
&-&\gamma[{\tilde \chi}_{s,c}^I(T_1)-{\tilde
\chi}_{s,c}^I(T)]\Bigr)H_s^2/3\nonumber\\
&+&(1-\delta)H_c^2\Bigl(\chi_{c,c}(T_1)-\chi_{c,c}(T)\Bigr)/6,
\label{HCI}
\end{eqnarray} where $T_1\approx 3.8$K is the
transition temperature at $H=0$. The apparent linearity of this
transition line thus follows from an approximate linear decrease
of $\chi_{c,c}(T)$ in the range 3.8K$<T<$6K, which also results in
a linear dependence of ${\tilde \chi}_{s,c}^C(T)$ and ${\tilde
\chi}_{s,c}^I(T)$.

To estimate the slope of $H_{\rm CI}(T)$, we need detailed
information on the temperature dependence of the various
individual susceptibilities. Experimentally, we have only
information on the total susceptibility, $\chi_c \approx \chi_c^I
\approx \chi_c^C$. Therefore, we are not able to make any
quantitative predictions based on the observed slope.

%To get some estimate, we start from the simple limit $J_{\rm
%av}=0$. In this limit, the susceptibilities of the spine spins do
%not depend on $T$, and we have $(\partial/\partial T)\chi_c
%\approx (\partial/\partial T)\chi_{c,c}/3$, for both phases. From
%the magnetization data we deduce $(\partial/\partial T)\chi_c
%\approx -0.005\mu_B/$(K~T~Ni), hence $(\partial/\partial
%T)\chi_{c,c} \approx -0.015\mu_B/$(K~T~Ni). Translating 1T$\approx
%1.3$K, and using the experimental $M_0 \approx 0.05 \mu_B/$Ni,
%this gives $(\partial/\partial T)\chi_{c,c}/M_0 \approx -0.08/{\rm
%K}^2$. From the phase diagram, we read $\Delta H_{C,I}/\Delta T
%\approx 1.5$T/K$\approx 1.9$. Using Eq. (\ref{HCI}), and noting
%that $(1-\delta)$ is of order 1, we conclude that $H_c$ is of
%order 12K, and therefore that $J_{\rm pd}$ is of order 3K. Since
%our direct magnetization estimate for this energy was very small,
%we must conclude that the uncertainties of these estimates are
%large. However, they all indicate that the order of magnitude for the
%PD energies to be around 1K.

\subsubsection{Other C--LTI phase boundaries}

For fields along (010) or (100), there is no linear term $HM_0$ in
the analog of Eq. (\ref{EC}). We also need to replace the relevant
susceptibilities, according to the direction of the field. For a
field in the $\alpha$-direction, the phase boundary is thus given
by
\begin{eqnarray}
&&\frac{1}{2}(\chi_\alpha^C-\chi_\alpha^I)H_{\rm CI}^2\nonumber\\
&=&\Bigl({\tilde
\chi}_{s,\alpha}^C(T_1)-{\tilde
\chi}_{s,\alpha}^C(T)-\gamma[{\tilde
\chi}_{s,\alpha}^I(T_1)-{\tilde
\chi}_{s,\alpha}^I(T)]\Bigr)H_s^2/3\nonumber\\
&+&(1-\delta)H_c^2\Bigl(\chi_{c,\alpha}(T_1)-\chi_{c,\alpha}(T)\Bigr)/6,
\label{HCI1} \end{eqnarray} yielding a parabolic dependence of $T$
on the transition field $H_{\rm CI}$. As seen from Fig. \ref{chi},
$(\chi_a^C-\chi_a^I)$ is larger and of opposite sign to
$(\chi_b^C-\chi_b^I)$, explaining the shapes of the two other
phase boundaries.

\subsection{Temperature and orientation dependence of the susceptibility}

We now return to Fig.~\ref{MvH}. To a good approximation, all the
curves in this figure can be described by straight lines, as
predicted in Eq. (\ref{MvsH}). The intercept $M_0$ is zero, except
for the high field data with ${\bf H} \parallel (001)$. In the
latter case, $M_0$ extrapolates to a value which seems to be
temperature independent.\par

The data in Fig.~\ref{MvH} exhibit two types of transitions: for
the field along (100) (upper panel), there is a transition from
the C phase (low $H$) to the LTI phase (high $H$). The only change
at this transition is a discontinuous increase in $\chi_{100}$.
This agrees with our expectations: assuming that the change comes
mainly from the ordered spine spins, this susceptibility is
longitudinal in the C phase, and has some contributions from the
transverse components in the LTI phase. In both phases, the
spontaneous moment $M_0$ along (100) vanishes.\par

For a field along (001) (lowest panel), there is a transition from
the LTI phase (low $H$) to the C phase, where, $\chi_{001}$ remains
almost $H$-independent, but $M_0$ exhibits a jump. The main
gain in energy comes from the spontaneous uniform magnetization.
In addition, in the C phase, $\chi_{001}$ is fully transverse,
whereas in the LTI phase it contains at least some longitudinal
component. However, this difference seems too small to be
observed.\par

The susceptibilities $\chi_\alpha$ all depend on temperature, with
$\chi_{010}$ and $\chi_{001}$ decreasing as $T$ increases, while
$\chi_{100}$ increases with increasing $T$. To analyze these
results quantitatively, we must use Eq. (\ref{chial}), which
requires some assumptions on the separate susceptibilities on the
spine and on the cross-tie sites. Qualitatively, it is reasonable
to guess that the increase in $\chi$ with decreasing $T$ for $H$
along (001) and (010) comes from the increase in the corresponding
(possibly Curie-like) susceptibilities of the cross-tie spins,
which are disordered. For fields along (100), the observed
decrease in $\chi$ with decreasing $T$ is probably due to the
strong decrease in the longitudinal spine susceptibility. However,
a full quantitative analysis requires more experimental information
than is currently available.

\subsection{Conjectured parameters}

\begin{figure}[ht]
\begin{center}
\includegraphics[width=9cm]{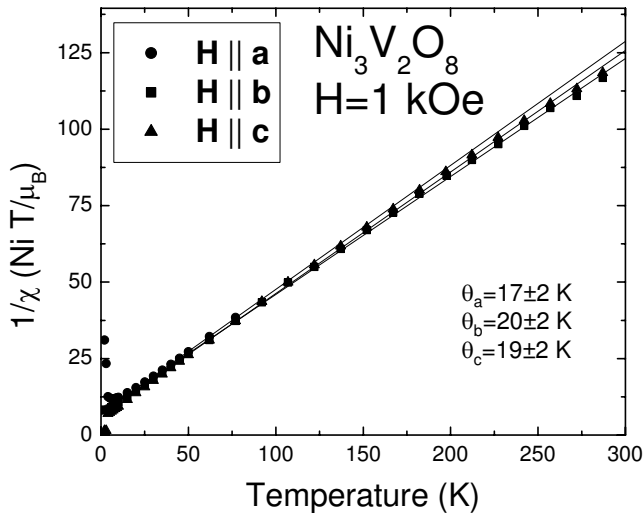}
\caption{\label{CURIEFIG} Determination of the Curie-Weiss temperatures
$\Theta_\alpha$ from the high-temperature susceptibility
data.\protect{\cite{Gavin}}}
\end{center}
\end{figure}

The most robust conclusion from our experiments and analysis is
Eq.  (\ref{J1J2}), for the ratio $J_{1a}/J_{2a}$. Additional
information concerning these exchange energies may be obtained
from the high temperature susceptibility tensor, whose components
are given asymptotically at high temperature as
\begin{eqnarray}
\chi_{\alpha \alpha} \approx C_\alpha/(T+\Theta_\alpha) \
,\label{newCurie}
\end{eqnarray}
where $\alpha$ labels the Cartesian component.  ($\chi_{\alpha \beta}$
must vanish for $\alpha \not= \beta$ for Cmca crystal structure.)
For NVO the Hamiltonian is of the form
\begin{eqnarray}
{\cal H} &=& - \sum_i \sum_\alpha g_\alpha (i) \mu_B S_\alpha(i) H_\alpha
\nonumber \\
&& + {1 \over 2} \sum_{k,l;\alpha , \gamma}
M_{\alpha \gamma} (k,l) S_\alpha (k) S_\gamma (l) \ ,
\end{eqnarray}
where $M_{\alpha \gamma}(i,i)$ is symmetric.  Generalizing Eq.
(\ref{Curie}), we find\cite{THEORY} that
\begin{eqnarray}
C_\alpha = [(\mu_B)^2 /3][8g_\alpha (s)^2 + 4 g_\alpha (c)^2 ] \ ,
\end{eqnarray}
where s denotes spine and c cross-tie, and
\begin{eqnarray}
C_\alpha \Theta_\alpha &=& {4 \mu_B^2 \over 9} \sum_{i=1}^6 \sum_{j \not=i}
g_\alpha (i) g_\alpha (j) M_{\alpha \alpha} (i,j) \nonumber \\
&& + {\mu_B^2 \over 3} \sum_i g_\alpha (i)^2 m_{\alpha \alpha}(i,i) \ ,
\end{eqnarray}
where the sums count all the interactions in one primitive unit cell.
If ${\bf g}(s) = {\bf g}(c)$ ($g \approx 2.3$\cite{BLEAN}), then
\begin{eqnarray}
C_\alpha = 4 \mu_B^2 g_\alpha^2
\end{eqnarray}
and
\begin{eqnarray}
\Theta_\alpha &=& {1 \over 9} \sum_{i=1}^6
\sum_{j\not = i} M_{\alpha \alpha } (i,j) +
{1 \over 12} \sum_i M_{\alpha \alpha } (i,i) \ .
\end{eqnarray}
If, for instance, the exchange interactions are isotropic, then
\begin{eqnarray}
\Theta_a = {8 \over 9} (J_1+J_2+J_b+J_c) + {16 \over 9} J_{\rm
av} - {2 \over 9} K \ ,
\end{eqnarray}
and the other components are
\begin{eqnarray}
\Theta_{b,c} = {8 \over 9} (J_1+J_2+J_b+J_c) + {16 \over 9} J_{\rm av} +
{1 \over 9} K \ ,
\end{eqnarray}
where $J_{\rm av}$ is the isotropic nn spine-cross-tie
interactions and the anisotropy constant $K$ is defined by
\begin{eqnarray}
{\cal H}_{\rm A}(i) &=& -K ( S_{ix}^2-2/3) \nonumber \\
&=& - {K\over 3} [ 2S_{ix}^2- S_{iy}^2 - S_{iz}^2].
\end{eqnarray}
Here we only attributed anisotropy to the spines. Now we consider
what values can be obtained from the data. For high temperatures,
a fit of the susceptibilities to Eq. (\ref{newCurie}), as shown
in Fig. \ref{CURIEFIG},  yields
$\Theta_{a,b,c}=17\pm 2$K, $20\pm 2$K, and $19\pm 2$K, respectively.
These values are not really consistent with an easy axis anisotropy.
Rather they indicate an
easy plane anisotropy with a smaller anisotropy which favors ${\bf
a}$ over ${\bf c}$. Using just the average value of 19K, we obtain
the constraint
\begin{eqnarray}
{8 \over 9} (J_1+J_2+J_b+J_c) + {16 \over 9} J_{\rm av} \approx
19{\rm K}.
\end{eqnarray}

We next estimate the anisotropy $K$. Figure \ref{PHASETZ} implies
that the LTI phase appears between the HTI and the C phases only
when $K/J_1 \sim 1/2$. This value is also roughly consistent with
the anisotropy of the susceptibility.

Given these constraints, we propose a possible set of exchange
parameters:
\begin{eqnarray}
J_1 &=& 10 \ , \ \ J_2= 4\ ,\ \ J_{\rm av} = 3\ , \nonumber \\
J_b &=& J_c=2\ ,\ \ {\rm and}\ \  K=5 \ ,
\label{EST}
\end{eqnarray}
all in units of K. The uncertainties are probably of order 50\%.
Substituting these values into Eq. (\ref{TPHEQ}) then yields the
mean field transition temperature $T_{\rm PH} \approx 20$K, which is
about 2.2 times the actual transition temperature. This reduction
from the mean field estimate must result from fluctuations. In ideal
Kagom\'{e} geometry one might think that $J_{\rm av}$ and $J_1$
would be identical. However, in NVO the Ni-O-Ni bond
angles\cite{Rogado} for $J_1$ ($95.0^{\rm o}$ and $90.4^{\rm o}$)
are further from $90^{\rm o}$ than those for $J_{\rm av}$
($90.3^{\rm o}$ and $91.5^{\rm o}$), which explains why $J_1$ is
larger than $J_{\rm av}$. Our estimates for the DM and PD parameters
were already given in Secs.~\ref{secDM} and \ref{secPD}.

\section{Conclusions}

In this paper we have presented a comprehensive investigation of the
magnetic phase diagram of NVO. Here we summarize our conclusions and
call attention to a number of topics for future research. Our
conclusions are

1) The magnetic structures of the HTI, LTI and C magnetic phases of
NVO have been closely determined. However, there are still some
uncertainties in the structure: some of the complex phases of the
various complex order parameters are not determined with precision.
In addition, for the LTI phase neutron diffraction data, on its own,
does not unambiguously identify whether the additional
representation which is characteristic of the LTI phase is
$\Gamma_1$ or $\Gamma_2$. However, the symmetry analysis of the
spontaneous polarization (see Ref. \onlinecite{FERRO}) indicates
that the correct choice is $\Gamma_1$.

2) We showed that a model having nearly isotropic interactions
between nn and nnn on a spine and with single ion anisotropy can
explain qualitatively the observed structures. (See Fig.
\ref{PHASETZ}.) In particular, the experimental determination of
the incommensurate wavevector accurately determines the ratio
$J_1/J_2$ of nearest- to next-nearest-neighbor interactions on a
spine.

3) Using the experimental values for the P to HTI transition
temperature and the Curie-Weiss temperatures (of the high
temperature uniform susceptibility), we obtain the estimates $J_1
\sim 10$K and $J_2 \sim 4$K.

4) From the appearance of small off-axis components of the
magnetization in the HTI phase we obtain estimates in the range $|D|
\sim 0.5$K for the Dzyaloshinskii-Moriya interactions between
adjacent Ni spine spins.

5) From the appearance of a weak FM moment in the commensurate AF
phases (and the appearance of transverse spin components in the HTI
phase) we conclude the existence of either anisotropic symmetric
exchange interactions or, more probably, Dzyaloshinskii-Moriya
antisymmetric spine-cross-tie exchange interactions.  Both
interactions are permitted by crystal symmetry and we give very
crude estimates of their values.

6) We have provided a qualitative explanation for the shape of the
phase boundaries between the various ordered and paramagnetic
phases. In principle the shape of these phase boundaries can be used
to deduce additional microscopic interactions. However, there are
currently too many unknown parameters to allow this program to be
carried out.

This work suggests several fruitful lines of future research, some
of which are ongoing. Since the nn and nnn interactions along the
spine are of the same order of magnitude, it would be interesting to
alter the geometry of the coordinating oxygen ions. This might be
done by applying pressure, or possibly uniaxial stress. Changing the
ratio  of these interactions would profoundly affect the magnetic
properties of NVO. It is even possible that the magneto-electric
properties could be grossly affected. Such experiments could greatly
expand the emerging understanding of the magnetoelectric behavior of
frustrated quantum magnets.

\begin{acknowledgments}
Work at Johns Hopkins University was supported by the National
Science Foundation through Grant No. DMR-0306940. Part of the work
was supported by the Israel US Binational Science Foundation under
grant number 2000073. The work at SPINS was supported by NSF
through DMR-9986442.
\end{acknowledgments}

\appendix

\section{Magnetic group theory analysis}
\label{Subapp_group_theory} Using the space group symmetry of the
crystal structure we identify allowed basis vectors for a magnetic
structure with the observed wave-vectors ${\bf v}$. This is done by
determining the irreps and eigenvectors of the
little group $G_{{\bf v}}$ of symmetry operations that leave the
wave-vectors ${\bf v}$ invariant. The eigenvectors $\phi^{\lambda}$
of the $\lambda$-th irreps $\Gamma^{\lambda}$ were determined using
the projector method.\cite{Heine} They are given by
\begin{equation}
    \phi^{\lambda}=\sum_g \chi^{\lambda}(g) g(\phi)\, ,
    \label{projecter}
\end{equation}where $g$ is an element of the little group and
$\phi$ is any vector of the order parameter space.
$\chi^{\lambda}(g)$ is character of symmetry element g in the
representation $\Gamma^{\lambda}$.\par

The symmetry elements of the \textit{Cmca} space group of NVO are
given in Table \ref{TablePUC}.  Note that this space group is
nonsymmorphic because some of the group elements $\{{\cal O}|{\bf
a}\}$ consist of a reflection or rotation ${\cal O}$ and require a
translation ${\bf a}$ equal to half a direct lattice vector.

\subsection{Commensurate structure}
The ordering wave-vector ${\bf v}=(0,0,0)$ is invariant under all
these operations so that the little group $G_{\bf v}$ contains all
the elements of the space group. The group consists of $8$ different
classes and therefore has $8$ irreps, all of which are one dimensional.
We determined the classes and the character table of this group and
these are shown in Table~\ref{comcharactertable}.\par

\begin{table}
\begin{tabular}{cc|ccccccccc}\hline\hline
& &$1$ &  $2_b$ & $2_a$ & $2_c$ & $\overline{1}$ & $m_{ac}$ &
$m_{bc}$ & $m_{ab}$\\
\hline
& $\Gamma^1$ & 1 &  1 &  1 &  1 &  1 &  1 &  1 &  1 &\\
& $\Gamma^2$ & 1 &  1 &  1 &  1 & -1 & -1 & -1 & -1 &\\
& $\Gamma^3$ & 1 &  1 & -1 & -1 &  1 &  1 & -1 & -1 &\\
& $\Gamma^4$ & 1 &  1 & -1 & -1 & -1 & -1 &  1 &  1 &\\
& $\Gamma^5$ & 1 & -1 &  1 & -1 &  1 & -1 &  1 & -1 &\\
& $\Gamma^6$ & 1 & -1 &  1 & -1 & -1 &  1 & -1 &  1 &\\
& $\Gamma^7$ & 1 & -1 & -1 &  1 &  1 & -1 & -1 &  1 &\\
& $\Gamma^8$ & 1 & -1 & -1 &  1 & -1 &  1 &  1 & -1 &\\
\hline\hline
\end{tabular}
\caption{\label{comcharactertable}Irreducible representation of the
group $G_{\bf v}$ for the commensurate magnetic structure with ${\bf
v}=(0,0,0)$.}
\end{table}

The eigenvectors were calculated using the projector method
(See Eq.~(\ref{projecter}).)
and are given in Table~\ref{TableIrreducibleRepCom}.

%\begin{widetext}

\begin{table}
\begin{scriptsize}
\begin{tabular}{c|c|c|c|c|c|c|c|c}\hline\hline
&$\psi_1$ & $\psi_2$ & $\psi_3$ & $\psi_4$ & $\psi_5$ & $\psi_6$ &
$\psi_7$ & $\psi_8$\vspace{0.05cm}\\ \hline

${\bf m}_{s1}$&$\begin{array}{c}0\\m^b_s\\0\end{array}$ &
$\begin{array}{c}0\\m^b_s\\0\end{array}$ &
$\begin{array}{c}0\\m^b_s\\0\end{array}$ &
$\begin{array}{c}0\\m^b_s\\0\end{array}$ &
$\begin{array}{c}m^a_s\\0\\m^c_s\end{array}$ &
$\begin{array}{c}m^a_s\\0\\m^c_s\end{array}$ &
$\begin{array}{c}m^a_s\\0\\m^c_s\end{array}$ &
$\begin{array}{c}m^a_s\\0\\m^c_s\end{array}$\\ \hline

${\bf m}_{s2}$&$\begin{array}{c}0\\-m^b_s\\0\end{array}$ &
$\begin{array}{c}0\\-m^b_s\\0\end{array}$ &
$\begin{array}{c}0\\m^b_s\\0\end{array}$ &
$\begin{array}{c}0\\m^b_s\\0\end{array}$ &
$\begin{array}{c}m^a_s\\0\\-m^c_s\end{array}$ &
$\begin{array}{c}m^a_s\\0\\-m^c_s\end{array}$ &
$\begin{array}{c}-m^a_s\\0\\m^c_s\end{array}$ &
$\begin{array}{c}-m^a_s\\0\\m^c_s\end{array}$\\ \hline

${\bf m}_{s3}$&$\begin{array}{c}0\\m^b_s\\0\end{array}$ &
$\begin{array}{c}0\\-m^b_s\\0\end{array}$ &
$\begin{array}{c}0\\m^b_s\\0\end{array}$ &
$\begin{array}{c}0\\-m^b_s\\0\end{array}$ &
$\begin{array}{c}m^a_s\\0\\m^c_s\end{array}$ &
$\begin{array}{c}-m^a_s\\0\\-m^c_s\end{array}$ &
$\begin{array}{c}m^a_s\\0\\m^c_s\end{array}$ &
$\begin{array}{c}-m^a_s\\0\\-m^c_s\end{array}$\\ \hline

${\bf m}_{s4}$&$\begin{array}{c}0\\-m^b_s\\0\end{array}$ &
$\begin{array}{c}0\\m^b_s\\0\end{array}$ &
$\begin{array}{c}0\\m^b_s\\0\end{array}$ &
$\begin{array}{c}0\\-m^b_s\\0\end{array}$ &
$\begin{array}{c}m^a_s\\0\\-m^c_s\end{array}$ &
$\begin{array}{c}-m^a_s\\0\\m^c_s\end{array}$ &
$\begin{array}{c}-m^a_s\\0\\m^c_s\end{array}$ &
$\begin{array}{c}m^a_s\\0\\-m^c_s\end{array}$\\ \hline

${\bf m}_{c1}$&$\begin{array}{c}m^a_c\\0\\0\end{array}$ & &
$\begin{array}{c}0\\m^b_c\\m^c_c\end{array}$ & &
$\begin{array}{c}m^a_c\\0\\0\end{array}$ & &
$\begin{array}{c}0\\m^b_c\\m^c_c\end{array}$\\ \hline

${\bf m}_{c2}$&$\begin{array}{c}-m^a_c\\0\\0\end{array}$ & &
$\begin{array}{c}0\\m^b_c\\-m^c_c\end{array}$ & &
$\begin{array}{c}m^a_c\\0\\0\end{array}$ & &
$\begin{array}{c}0\\-m^b_c\\m^c_c\end{array}$\\ \hline \hline
\end{tabular}
\caption{\label{TableIrreducibleRepCom}Irreducible representations
for the commensurate phase described with ${\bf v}=(0,0,0)$ for both
the ${\rm Ni_s}$ and ${\rm Ni_c}$ sites. The components of the
vector correspond to the spin component on the Ni sites in the order
given in Table~\protect\ref{NiLattice}.}
\end{scriptsize}
\end{table}

%\end{widetext}

\subsection{Incommensurate structure}
For the ordering wave-vector ${\bf v}=(q,0,0)$, the little group
$G_{\bf v}$ contains the following four elements of the space group:
\begin{equation}
    \{ 1,\;2_x,\;m_{xy},\;m_{xz} \}\, ,
\end{equation}The group consists of $4$ different classes and
therefore has $4$ one-dimensional irreps. We determined the
classes and the character table of this group and these are shown
in Table~\ref{incomcharactertable}.\par

\begin{table}
\begin{tabular}{cc|ccccc}\hline\hline
& &$1$ &  $2_a$ & $m_{ac}$ & $m_{ab}$ & \\ \hline
& $\Gamma^1$ & 1 &  1 &  1 &  1 &\\
& $\Gamma^2$ & 1 &  1 & $-1$ & $-1$ &\\
& $\Gamma^3$ & 1 & $-1$ &  $1$ & $-1$ &\\
& $\Gamma^4$ & 1 & $-1$ & $-1$ &  $1$ &\\ \hline\hline
\end{tabular}
\caption{\label{incomcharactertable}Irreducible representation of
the group $G_{\bf v}$ for the incommensurate magnetic structure with
${\bf v}=(q,0,0)$.}
\end{table}

The eigenvectors were calculated using the projector method using
Eq.~ (\ref{projecter}) and are given in
Table~\ref{TableIrreducibleRepInCom}.

\begin{table}[ht]
\begin{scriptsize}
\begin{tabular}{c|c|c|c|c|c|c|c|c|ccc}\hline\hline
&$\psi_1$ & $\psi_2$ & $\psi_3$ & $\psi_4$ \vspace{0.05cm}\\
\hline

${\bf m}_s^1$&$\begin{array}{c}m^a_s\\m^b_s\\m^c_s\end{array}$ &
$\begin{array}{c}m^a_s\\m^b_s\\m^c_s\end{array}$&
$\begin{array}{c}m^a_s\\m^b_s\\m^c_s\end{array}$&
$\begin{array}{c}m^a_s\\m^b_s\\m^c_s\end{array}$\\ \hline

${\bf m}_s^2$&$\begin{array}{c}m^a_s\\-m^b_s\\-m^c_s\end{array}$&
$\begin{array}{c}m^a_s\\-m^b_s\\-m^c_s\end{array}$&
$\begin{array}{c}-m^a_s\\m^b_s\\m^c_s\end{array}$&
$\begin{array}{c}-m^a_s\\m^b_s\\m^c_s\end{array}$\\ \hline

${\bf m}_s^3$ & $\begin{array}{c} - m^a_s\\ m^b_s\\
- m^c_s\end{array}$ & $\begin{array}{c} m^a_s\\ -m^b_s\\
m^c_s \end{array}$ & $\begin{array}{c}-m^a_s\\
m^b_s\\-m^c_s\end{array}$&
$\begin{array}{c} m^a_s\\ - m^b_s \\ m^c_s\end{array}$\\ \hline

${\bf m}_s^4$ & $\begin{array}{c} - m^a_s \\ -m^b_s \\
m^c_s\end{array}$ & $\begin{array}{c} m^a_s\\ m^b_s\\
-m^c_s \end{array}$ & $\begin{array}{c} m^a_s\\ m^b_s\\
-m^c_s \end{array}$ & $\begin{array}{c} -m^a_s \\ -m^b_s\\
m^c_s\end{array}$\\ \hline

${\bf m}_c^1$&$\begin{array}{c}m^a_c\\0\\0\end{array}$&
$\begin{array}{c}m^a_c\\0\\0\end{array}$&
$\begin{array}{c}0\\m^b_c\\m^c_c\end{array}$&
$\begin{array}{c}0\\m^b_c\\m^c_c\end{array}$\\ \hline

${\bf m}_c^2$ & $\begin{array}{c} -m^a_c\\ 0\\0\end{array}$&
$\begin{array}{c}  m^a_c\\ 0\\0\end{array}$&
$\begin{array}{c} 0\\ m^b_c\\ -m^c_c\end{array}$&
$\begin{array}{c} 0\\ -m^b_c \\ m^c_c\ \end{array}$\\
\hline \hline
\end{tabular}\caption{\label{TableIrreducibleRepInCom}Irreducible
representations for the incommensurate phase associated with ${\bf
v}=(q,0,0)$ for the ${\rm Ni_s}$ and ${\rm Ni_c}$ sites.  The components
of the vector correspond to the spin component on the Ni sites in the
order given in Table~\protect\ref{NiLattice}.}
\end{scriptsize}
\end{table}

As explained in the text, we replace the variables in the above table
by symmetry adapted coordinates, so that we parametrize the eigenvectors
as in Table \ref{SYMAD}.

To fit to data one assumes a single irrep
and then optimizes with respect to the choice of the
complex-valued amplitudes $m_s^a$, $m_s^b$, {\it etc.}
For the HTI phase one selects the representation which
best fits the data for the optimized values of the
amplitudes.  Alternatively, one can introduce the symmetry
adapted coordinates as in Eqs. (\ref{SYMEQ})
and accept the conclusion from Landau theory that these
symmetry adapted coordinates all have the same
complex phase.  These symmetry adapted coordinates are
constructed by the methods leading to Eqs. (\ref{SYMEQ})
and are given in Table \ref{SYMAD}.

\begin{table}
\begin{scriptsize}
\begin{tabular}{c|c|c|c|c|c|c|c|c|ccc}\hline\hline
&$\psi_1$ & $\psi_2$ & $\psi_3$ & $\psi_4$ \vspace{0.05cm}\\
\hline

${\bf m}_s^1$&$\begin{array}{c}i\tilde m_{sa} \\ \tilde m_{sb}\\
i\tilde m_{sc} \end{array}$ &
$\begin{array}{c}\tilde m_{sa} \\ i\tilde m_{sb} \\ \tilde m_{sc}
\end{array}$&
$\begin{array}{c} i\tilde m_{sa} \\ \tilde m_{sb} \\ i\tilde m_{sc}
\end{array}$&
$\begin{array}{c} \tilde m_{sa} \\ i\tilde m_{sb} \\ \tilde m_{sc}
\end{array}$\\ \hline

${\bf m}_s^2$&$\begin{array}{c} i\tilde m_{sa} \\ -\tilde m_{sb}
\\ -i\tilde m_{sc} \end{array}$&
$\begin{array}{c} \tilde m_{sa} \\ -i\tilde m_{sb} \\ -\tilde m_{sc}
\end{array}$&
$\begin{array}{c} -i\tilde m_{sa} \\ \tilde m_{sb} \\ i\tilde m_{sc}
\end{array}$&
$\begin{array}{c} -\tilde m_{sa} \\ i\tilde m_{sb} \\ \tilde m_{sc}
\end{array}$\\ \hline

${\bf m}_s^3$&$\begin{array}{c} - i \tilde m_{sa} \\
\tilde m_{sb} \\ - i \tilde m_{sc} \end{array}
$&$\begin{array}{c} \tilde m_{sa} \\ -i \tilde m_{sb} \\
\tilde m_{sc} \end{array}$&$
\begin{array}{c} -\tilde m_{sa} \\ i \tilde m_{sb} \\
-\tilde m_{sc} \end{array}$&
$\begin{array}{c} \tilde m_{sa} \\ -i \tilde m_{sb} \\
\tilde m_{sc} \end{array}$\\ \hline

${\bf m}_s^4$&$\begin{array}{c} - i \tilde m_{sa} \\
 - \tilde m_{sb} \\ i \tilde m_{sc} \end{array}
$&$\begin{array}{c} \tilde m_{sa} \\ i \tilde m_{sb} \\
 -\tilde m_{sc} \end{array}$&$
\begin{array}{c} \tilde m_{sa} \\ i \tilde m_{sb} \\
-\tilde m_{sc} \end{array}$&
$\begin{array}{c} -\tilde m_{sa} \\ -i \tilde m_{sb} \\
\tilde m_{sc} \end{array}$\\ \hline

${\bf m}_c^1$&$\begin{array}{c} \tilde m_{ca}\\0\\0\end{array}$&
$\begin{array}{c} \tilde m_{ca}\\0\\0\end{array}$&
$\begin{array}{c}0\\ \tilde m_{cb}\\ \tilde m_{cc} \end{array}$&
$\begin{array}{c}0\\ \tilde m_{cb}\\ \tilde m_{cc}\end{array}$\\ \hline

${\bf m}_c^2$&$\begin{array}{c}- \tilde m_{ca} \\ 0\\0\end{array}$&
$\begin{array}{c} \tilde m_{ca} \\ 0\\0\end{array}$&
$\begin{array}{c} 0\\ \tilde m_{cb} \\ -\tilde m_{cc} \end{array}$&
$\begin{array}{c} 0\\-\tilde m_{cb} \\ \tilde m_{cc}  \end{array}$\\
\hline \hline
\end{tabular}\caption{ \label{SYMAD}
Symmetry adapted coordinates which transform according to the irreps
for the incommensurate phase associated with ${\bf v}=(q,0,0)$ for
the Ni$_s$ and Ni$_c$ sites.  When more than one representation is active,
it is a reasonable approximation to allow all parameters
of representation $\Gamma_n$ to have the same phase factor $e^{i\phi_n}$.}
\end{scriptsize}
\end{table}

\section{Magnetic neutron scattering cross-section}
The integrated intensity of magnetic Bragg peaks is related to the
structure factor of the magnetic ordering through
\cite{Lovesey_book}
\begin{equation}\label{Eq_int}
    I({\bf Q})=\left(\frac{\gamma r_0}{2}\right)^2 N_m \frac{(2\pi)^3}{V_{m0}}
    \Phi \;R({\bf Q})\; |f({\bf Q})|^2 |{\bf F}_{\bot}({\bf Q})|^2\, ,
\end{equation} where $f({\bf Q})$ is the magnetic form factor for Ni$^{2+}$
ions.\cite{IntTables}  $\Phi$ is the flux of incident neutrons,
$N_{m}$ and $V_{m0}$ are the number and the volume of the magnetic
unit cells, $\gamma=-1.913$ and
$r_0=2.818\cdot10^{-15}\;\mathrm{m}$. $I({\bf Q})$ is the total
integrated intensity of a Bragg reflection when the sample is
rotated about the normal to the scattering plane and $R({\bf Q})$ is
a factor which takes into account the ${\bf Q}$-dependent
sensitivity of the spectrometer. For the powder experiment, $R({\bf
Q})$=$1/\sin(\theta(\bf Q))$. For the triple-axis experiment,
$R({\bf Q})$ was calculated using the Cooper-Nathans
approximation.\cite{Cooper_Nathans} ${\bf F}_{\bot}({\bf Q})$ is the
component of the magnetic structure factor perpendicular to the
scattering wave vector and is defined as
\begin{equation}\label{Eq_fperp}
    {\bf F}_{\bot}({\bf Q}) = {\bf F}({\bf Q}) -
    ({\bf F}({\bf Q}) \cdot {\bf \hat{Q}})\;{\bf \hat{Q}},
\end{equation}
where ${\bf \hat{Q}}= {\bf Q}/Q$.

The magnetic structure factor for a commensurate structure was
calculated using\vspace{0.5cm}
\begin{equation}\label{Eq_f}
    {\bf F}({\bf Q}) = \sum_{i} {\bf m}_i \exp (-{\rm i}\;
    {\bf Q} \cdot {\bf d}_i)
\end{equation}
where ${\bf d}_i$ are the positions of the Ni$^{2+}$-ions and the
sum is over the magnetic ions in the magnetic unit cell. The
magnetic dipole moments ${\bf m}_i$ are given by one of the irreps
given in Table~\ref{TableIrreducibleRepCom}. For incommensurate
structures the magnetic structure factor was calculated using
\begin{equation}\label{Eq_ff}
    {\bf F}({\bf Q}) = \sum_{i} \frac{1}{2}\;{\bf \Psi}^{\bf v}_i \exp (-{\rm i}\;
    {\bf Q} \cdot {\bf d}_i)\, ,
\end{equation}where the Fourier components of the structure,
$\Psi^{\bf v}_i$, are given by the basis vectors of the irreps
of the little group given in
Table~\ref{TableIrreducibleRepInCom}. The moment on site $i$ is
given by ${\bf m}_i$=$\frac{1}{2} ({\bf \Psi}^{\bf v} + ({\bf
\Psi}^{\bf v})^\star)$, so the factor $\frac{1}{2}$ in
Eq.~ (\ref{Eq_ff}) ensures that the scale factor for the basis vectors
of the irreps can be expressed in Bohr magnetons.

\section{Absolute magnitude of the ordered moment}
\label{Subapp_magnitude} The magnitude of the ordered moments were
determined by comparing the nuclear and the AF Bragg peak
intensities. The intensity of a nuclear Bragg peak is given as
\begin{equation}
    I({\bf Q})=N_{uc} \frac{(2\pi)^3}{V_{0}} \Phi
    \;R({\bf Q})\;|F_N({\bf Q})|^2\,
    \label{Eq_nuclint},
\end{equation}where $N_{uc}$ and $V_{0}$ are the number and the
volume of the unit cells. $F_N({\bf Q})$ is the nuclear structure
factor given as\cite{Squires}
\begin{equation}
    F_N({\bf Q})=\sum_{i}b_i \exp(-{\rm i}\, {\bf Q} \cdot {\bf d}_i)\,
    ,
\end{equation}where the sum runs over all elements in the nuclear
unit cell and $b_i$ is the bound coherent scattering
length\cite{IntTables} of atom $i$ in the unit cell. Since multiple
scattering and extinction corrections were important for the
strongest reflections of the single crystal, their intensities were
ignored for the normalization. The measured nuclear intensities gave
the overall scale factor $A_{\rm exp}=N_{uc} \frac{(2\pi)^3}{V_0}
\Phi$ in Eq~ (\ref{Eq_nuclint}). The magnitude of the ordered
magnetic moment was determined using $A_{\rm exp}$ and Eq.~
(\ref{Eq_int}), allowing then the determination of the ordered
moment for the various structures.\par

% ------------------------------------------------------------------------
\bibliographystyle{prsty}
\bibliography{references}
\end{document}